\documentclass[11pt,a4paper]{article}
\usepackage{jheppub_kim}
\usepackage{rotating} 
\usepackage{graphicx,epsfig}
\usepackage{amsmath}
\usepackage {amssymb}
\usepackage{subfigure}

\usepackage{relsize}

\def\beq{\begin{equation}}
\def\eeq{\end{equation}}
\def\bea{\begin{eqnarray}}
\def\eea{\end{eqnarray}}

\begin{document}

\title{Bounce and cyclic cosmology in weakly broken galileon theories}

\author[a]{Shreya Banerjee}

\author[b,c]{Emmanuel N. Saridakis}
 
 \affiliation[a]{Tata Institute of Fundamental 
Research, Homi Bhabha Road, Mumbai 400005, India}

\affiliation[b]{Instituto de F\'{\i}sica, Pontificia
Universidad de Cat\'olica de Valpara\'{\i}so, Casilla 4950,
Valpara\'{\i}so, Chile}

\affiliation[c]{CASPER, Physics Department, Baylor University, Waco, TX 76798-7310, USA}

\emailAdd{shreya.banerjee@tifr.res.in}

\emailAdd{Emmanuel$_-$Saridakis@baylor.edu}

\abstract{We investigate the bounce and cyclicity realization in the framework of 
weakly broken galileon theories. We study bouncing and cyclic solutions at the 
background level, reconstructing the potential and the galileon functions that can 
give rise to a given scale 
factor, and presenting analytical expressions for the bounce requirements. 
We proceed to a detailed investigation of the perturbations, which after crossing the 
bouncing point give rise to various observables, such as the scalar and tensor spectral 
indices and the tensor-to-scalar ratio. Although the scenario at hand 
shares the disadvantage of all bouncing models, namely that it provides a large 
tensor-to-scalar ratio, introducing an additional light scalar significantly 
reduces it through the kinetic amplification of the isocurvature 
fluctuations. 
}

\keywords{Modified gravity, Galileon symmetry, Bounce, 
Cyclic cosmology}

\maketitle

\section{Introduction}

Inflation is now considered to be a crucial part of the universe cosmological history  
\cite{inflation}, however the so called ``standard 
model of the universe'' still faces the problem of the initial singularity. Such a 
singularity is unavoidable if inflation is realized using a scalar field while the 
background spacetime is described by the standard Einstein action \cite{Borde:1993xh}. As 
a consequence, there has been a lot
of effort in resolving this problem through quantum gravity effects or effective field 
theory techniques.

A potential solution to the cosmological singularity problem may be provided by 
non-singular bouncing cosmologies \cite{Mukhanov:1991zn}. Such scenarios have been 
constructed through various approaches to modified gravity 
\cite{Nojiri:2006ri,Capozziello:2011et}, such as the Pre-Big-Bang \cite{Veneziano:1991ek} 
and the Ekpyrotic \cite{Khoury:2001wf,Khoury:2001bz} models, gravity actions with higher 
order corrections \cite{Tirtho1,Nojiri:2013ru}, $f(R)$ gravity 
\cite{Bamba:2013fha,Nojiri:2014zqa}, $f(T)$ 
gravity 
\cite{Cai:2011tc}, braneworld scenarios \cite{Shtanov:2002mb,Saridakis:2007cf}, 
non-relativistic gravity \cite{Cai:2009in,Saridakis:2009bv}, massive gravity  
\cite{Cai:2012ag}, Lagrange modified gravity \cite{Cai:2010zma}, loop quantum cosmology 
\cite{Bojowald:2001xe,Odintsov:2014gea,Odintsov:2015uca} 
or in the frame of a closed universe 
\cite{Martin:2003sf}. Non-singular bounces may be
alternatively investigated using effective field theory techniques, introducing matter 
fields violating the null energy condition 
\cite{Cai:2007qw,Cai:2009zp,Nojiri:2015fia}, or introduce non-conventional 
mixing 
terms \cite{Saridakis:2009jq,Saridakis:2009uk}. The extension of all the above bouncing 
scenarios is the (old) paradigm of cyclic cosmology \cite{tolman}, in which the universe 
experiences the periodic sequence of contractions and expansions, which has been rewaked 
the last years \cite{Steinhardt:2001st,Steinhardt:2002ih} since it brings different 
insights for
the origin of the observable universe 
\cite{Lidsey:2004ef,cyclic1,Nojiri:2011kd} (see \cite{Novello:2008ra} for a review). Such 
scenarios are also capable of explaining the 
scale invariant power spectrum 
\cite{Novello:2008ra,Finelli:2001sr} and moderate 
non-Gaussianities \cite{Cai:2009fn}. Hence, they are considered as a potential 
alternative to Big Bang cosmology. 

One very general class of gravitational modification are galileon theories 
\cite{Nicolis:2008in,Deffayet:2009wt,Deffayet:2009mn}, 
which are a re-discovery of Horndeski general scalar-tensor theory 
\cite{Horndeski:1974wa}, in which one introduces higher derivatives in the scalar-tensor 
action, with the requirement of maintaining the equations of motion second-ordered. In 
this 
formulation the Lagrangian is imposed to satisfy the Galilean symmetry 
$\phi\rightarrow\phi+b_\mu x^\mu$, with $b_\mu$ a constant, and an additional advantage 
is that the scalar field derivative self-couplings screen the deviations from General 
Relativity at high gradient regimes due to the Vainshtein mechanism 
\cite{Vainshtein:1972sx}, thus satisfying the solar system constraints. These features 
led galileon theories and their modifications to have an extensive application in 
cosmological frameworks. In particular, one can
study the late-time acceleration
\cite{Silva:2009km,Gannouji:2010au,DeFelice:2010pv,Tretyakov:2012zz,Leon:2012mt},
inflation
\cite{Creminelli:2010ba,Kobayashi:2010cm,Ohashi:2012wf} 
and non-Gaussianities
\cite{Mizuno:2010ag,Gao:2011qe,RenauxPetel:2011uk}, cosmological 
perturbations
\cite{Kobayashi:2009wr,DeFelice:2010as,Barreira:2012kk}, and use observational 
data to constrain various classes of galileon 
theories
\cite{Ali:2010gr,Iorio:2012pv,Appleby:2012ba}.
 
Recently, a model of weakly broken galileon symmetry appeared in the literature 
\cite{Pirtskhalava:2015nla}. In this construction the notion of weakly broken galileon 
invariance was introduced, which characterizes the unique class of gravitational 
couplings that maximally preserve the defining symmetry. Hence, the curved-space remnants
of the quantum properties of the galileon allow one to construct quasi de Sitter 
backgrounds that remain to a large extent insensitive to loop corrections 
\cite{Pirtskhalava:2015nla}.
 
In the present work, we are interested in investigating the bounce and cyclicity 
realization in the framework of weakly broken galileon theories. Although the bouncing 
realization has been shown to be possible in the context of usual galileon cosmology  
\cite{Qiu:2011cy,Easson:2011zy,Cai:2012va,Qiu:2013eoa}, we show that in the present 
weakly broken variance we have enhanced freedom to satisfy the relevant requirements.
The plan of the work is as follows: In Section \ref{galileon} we briefly review  
theories  with weakly broken galileon invariance, and we apply them in a cosmological
framework. In Section 
\ref{bouncecyclic} we investigate the realization of bouncing and cyclic solutions at 
the background level, reconstructing the corresponding potentials and the galileon 
functions.
 In Section \ref{perturbations} we analyze the perturbations of the scenario, and we 
study how they pass through the bouncing point, giving rise to various observables, such 
as the scalar and tensor spectral indices and the tensor-to-scalar ratio. Finally, in 
section \ref{Conclusions} we summarize our results.

\section{Cosmology with weakly broken galileon symmetry}
\label{galileon}

Let us briefly review theories with weakly broken galileon invariance following 
\cite{Pirtskhalava:2015nla}. Such constructions include a scalar field coupled to 
gravity, 
and form a subclass of Horndeski theories which only weakly breaks the galileon symmetry 
even in the presence of gravity. This property is achieved by suitably formulating these 
theories in order for the symmetry-breaking interaction terms in the Lagrangian to be 
suppressed. The advantage of this procedure is that the resulting field equations remain 
of second order, although the Lagrangian includes higher derivative interaction terms.

The action of this class of theories reads as \cite{Pirtskhalava:2015nla}
\begin{equation}
\label{fullaction}
S=\int d^4x \sqrt{-g}
\left[
\frac{1}{2}M_{pl}^2 R 
-\frac{1}{2}(\partial\phi)^2 
-V(\phi)+\sum_{I=2}^5 \mathcal{L}^{\rm WBG}_I +\dots\right]+S_m~,
\end{equation}
with $\phi$ the scalar field, $R$ the Ricci scalar, $M_{pl}$ the Planck mass, $S_m$ the 
matter-sector action, and where we have defined the operators $\mathcal{L}^{\rm WBG}_I$ 
to be given by the following subclass of the Horndeski terms:
\begin{align}
\label{hor1}
\mathcal{L}^{\rm WBG}_2&=\Lambda_2^4 ~G_2(X) ~,\\
\mathcal{L}^{\rm WBG}_3&=\frac{\Lambda_2^4}{\Lambda_3^3} ~G_3(X)[\Phi] ~,\\
\label{planck}
\mathcal{L}^{\rm WBG}_4&=\frac{\Lambda_2^8}{\Lambda_3^6}  ~ G_{4}(X) R+2 
\frac{\Lambda_2^4}{\Lambda_
3^6}  ~G_{4X}(X)\left( [\Phi]^2-[\Phi^2] \right)~,\\
\label{hor4}
\mathcal{L}^{\rm WBG}_5&=\frac{\Lambda_2^8}{\Lambda_3^9} 
~G_{5}(X)G_{\mu\nu}\Phi^{\mu\nu}-\frac{\Lambda_2^4}{3 \Lambda_3^9} ~ 
G_{5X}(X)\left([\Phi]^3-3[\Phi][\Phi^2]+2[\Phi^3] \right) ~.
\end{align}
In the above expressions $G_I$ are arbitrary dimensionless functions of the dimensionless 
variable 
\begin{equation}
X\equiv - \frac{1}{\Lambda^4_2} g^{\mu\nu}\partial_\mu\phi\partial_\nu\phi ~,  
\label{X_def}
\end{equation}
and we have used the subscript ``$X$'' to denote differentiation with respect to this 
variable, while $G_{\mu\nu}$ is the Einstein tensor. Furthermore, we have introduced the 
compact notation \cite{Pirtskhalava:2015nla}
\begin{eqnarray}
&&[\Phi]\equiv g^{\mu\nu}\nabla_\mu\nabla_\nu\phi\nonumber\\
&&[\Phi^2]
\equiv \nabla^\mu\nabla_\nu\phi\nabla^\nu\nabla_\mu\phi\nonumber\\
&&\ \ \ \ \ \ \cdots.
\end{eqnarray}
Additionally, the parameter $\Lambda_{3}$ marks the scale suppressing
the invariant galileon interactions, while the parameter $\Lambda_{2}=(M_{
pl}\Lambda_{3}^{3})^{1/4}$, with $\Lambda_{3}\ll \Lambda_{2}$, marks the significantly 
higher scale suppressing the quantum-mechanically generated single-derivative
operators \cite{Pirtskhalava:2015nla}. Obviously, in the limit where both 
$\Lambda_{2}$,$\Lambda_{3}$ go to $M_{pl}$, weakly broken galileon invariance 
disappears, and the above theories become the usual covariant galileon ones.
Note that in action 
(\ref{fullaction}) one can 
consider a potential $V(\phi)$, which is the only term that breaks the scalar shift 
symmetry, which is otherwise exact even in
curved space.

Let us now apply the above theories in a cosmological framework. In particular, we 
consider a flat 
Friedmann-Robertson-Walker (FRW) spacetime metric of the form
\begin{equation}
ds^{2}=-dt^{2}+a(t)^{2} \delta_{ij}dx^{i}dx^{j},
\end{equation}
where $a(t)$ is the scale factor. For this metric, the metric field equations 
derived from 
action (\ref{fullaction}) become the two Friedmann equations \cite{Pirtskhalava:2015nla}
\begin{eqnarray}
\label{fried1}
&&\!\!\!\!\!\!\!\!\!\!\!\!\!\!\!\!
3M_{pl}^2 H^2=\rho_m+ \ V 
+\Lambda_2^4X\bigg[\frac{1}{2}-\frac{G_2}{X}+2G_{2X}-6ZG_{3X}-6Z^2\left(\frac{G_{4}}
{X^2}-4\frac{G_{4X}}{X} -4G_{4XX} \right)\nonumber
\\&&
\ \ \ \ \ \ \,
   +2Z^3\left(5\frac{G_{5X}}{X}+2G_{5XX}\right)\bigg ]~, \\
\label{fried2}
&&\!\!\!\!\!\!\!\!\!\!\!\!\!\!\!\!
M_{pl}^2\dot H = -\frac{\Lambda_2^4 X F +M_{pl}\ddot \phi(X G_{3X}-4 
ZG_{4X}-8ZXG_{4XX}-3 
Z^2 G_{
5X}-2Z^2XG_{5XX})}{1+2 G_{4}-4X 
G_{4X}-2ZXG_{5X}}\nonumber\\
&&  \ \ \ \  -\frac{\rho_m}{2} -\frac{p_m}{2},
\end{eqnarray}
with $\rho_m$ and $p_m$ the energy density and pressure of the matter sector, 
assumed to correspond to a perfect fluid, and 
where $H=\dot{a}/a$ is the Hubble 
parameter and a dot denotes differentiation with respect to $t$. In the above expressions 
we have defined the function
\begin{equation}
\label{f}
F(X,Z) = \frac{1}{2}+G_{2X}-3 Z G_{3X}+6 Z^2 \left(\frac{G_{4X}}{X}+2 G_{4XX}\right)+ 
Z^3\left(3\frac{G_{5X}}{X}
+2 G_{5XX}\right)~,
\end{equation}
with the variable $Z$ defined as 
\beq
\label{Z_def1}
 Z \equiv \frac{ H \dot\phi}{\Lambda_3^3}~. 
\eeq
Additionally, the equation of motion for the scalar field becomes  
\cite{Pirtskhalava:2015nla}
\begin{equation}
\label{scalareom}
\frac{1}{a^3}\frac{d}{dt}\left[2a^3\dot{\phi}\,F(X,Z) \right]=-\frac{dV}{d\phi}.
\end{equation}
Finally, note that according to definition (\ref{X_def}), in FRW geometry we have  
$X=  \dot{\phi}^2/\Lambda^4_2$.  Lastly, note that the above equations close considering 
the matter conservation equation  
\begin{equation}
\label{mattconserv}
\dot{\rho}_m+3H(\rho_m+p_m)=0.
\end{equation}

\section{Background bouncing and cyclic solutions}
\label{bouncecyclic}

In this section we are interested in investigating the bounce and cyclicity realization 
in cosmologies with weakly broken galileon invariance, at the background level. Let us 
first review the basic 
conditions for these realizations. An expanding universe is characterized by a positive 
Hubble parameter, while a contracting one by a negative $H$. Using the 
continuity equations we deduce that at the bounce and turnaround points $H=0$. However, 
at and around the bounce we must have $\dot H> 0$, while at and around the turnaround we 
obtain  $\dot H < 0$.

One can easily see that the above conditions cannot be fulfilled in 
the framework of general relativity, nevertheless they can be easily satisfied in the 
scenario at hand. In particular, observing the form of the two Friedmann equations 
(\ref{fried1}),(\ref{fried2}), along with the scalar-field equation (\ref{scalareom}), we 
conclude that for suitable choices of the free functions $G_I$ and of the scalar 
potential $V(\phi)$ one can acquire the necessary violation of the null energy condition 
and hence the satisfaction of the bouncing and cyclic conditions.

Let us make an important point concerning the bounce and cyclicity reconstruction. In 
principle, in the present scenario, one has the freedom to determine the free functions 
 $G_I$'s as well as the scalar potential $V(\phi)$. However, note that while in the case 
where the  $G_I$'s are all zero (i.e in the case of minimally-coupled general relativity) 
there is no potential that can drive a bounce, in the case of suitably chosen non-trivial 
 $G_I$'s a bounce can be realized either with a zero potential or with a suitably chosen 
non-zero potential. From these we deduce that the crucial ingredient of bounce and 
cyclicity realization is the Galileon functions  $G_I$'s and not the potential $V(\phi)$. 
This feature, along the fact that in Galileon construction shift symmetry plays a crucial 
role and thus a potential is absent, led the initial works of Galileon cosmology, and 
Galileon bouncing cosmology in particular, not to consider a scalar potential and focus on 
the special choice of the  $G_I$'s functions \cite{Qiu:2011cy,Easson:2011zy}. 
Nevertheless, since in the generalized Galileon theory (or in the point of view of 
Horndeski theory), which is the basis of the present work, a scalar potential is allowed, 
one has an additional free function to play with, and thus he can alleviate some tuning 
from the functions $G_I$'s. 

Hence, in the following subsections, for completeness, we will reconstruct bounce and 
cyclic cosmology reconstructing first the necessary scalar potential for a not so tuned 
choice of the functions $G_I$'s, and then reconstructing  $G_I$'s for a zero or non-zero 
given potential.

\subsection{Reconstruction of a bounce}
\label{reconstructionb}

Let us now present the bounce realization at the background level. Without loss of 
generality we consider a bouncing scale factor of the form  
\begin{equation}
a(t)=a_b(1+Bt^2)^{1/3},
\label{matterbounce}
\end{equation}
where $a_b$ is the scale factor value at the bounce, while 
$B$ is a positive parameter which determines how fast the
bounce takes place. In this case time varies between $-\infty $ and $+\infty $, 
with $t=0$ the bouncing point. Hence, since the scale factor is known we can 
straightforwardly find the forms of $H(t)$ and $\dot{H}(t)$ as 
\begin{eqnarray}
\label{Htt}
&&H(t)=\frac{2Bt}{3(1+Bt^2)}\\
&&\dot{H}(t)=\frac{2B}{3}\left[\frac{1-Bt^2}{(1+Bt^2)^2}\right].
\label{Hdottt}
\end{eqnarray}

As we discussed above, one can realize the above background bouncing solution either 
choosing (without tuning) the forms of all the functions $G_I$'s and suitably 
reconstruct the scalar potential $V(\phi)$, or choose a zero or a simple $V(\phi)$ and 
some of the $G_I$'s and suitably reconstruct the remaining $G_I$. In the following we 
investigate these two procedures separately.

\subsubsection{Reconstructing  $V(\phi)$}
\label{reconstructionb111}

Let us first study the case where the ansatzes for the functions $G_I$'s are considered 
by hand, without any particular form of tuning. According to the discussion in 
\cite{Pirtskhalava:2015nla}, in theories with weakly broken galileon invariance the 
functions $G_2$ and $G_4 $ should be assumed to start at least quadratic in $X$. Hence, 
the simplest class of models with weakly broken 
galileon symmetry would be
\begin{equation}
G_{2}=G_{4}=X^2;\ \ \ G_3=X  ; \ \ \ G_5=0.
\label{ansGI1}
\end{equation}
Inserting (\ref{matterbounce}) and (\ref{ansGI1}) into the Friedmann equations 
(\ref{fried1}),(\ref{fried2}) we obtain 
\begin{eqnarray}
&&\!\!\!\!\!\!\!\!\!\!\!
3M_{pl}^{2}H(t)^{2}=\rho_m(t)+V(\phi(t))+\dot{\phi}(t)^2\left[\frac{1}{2}+\frac{3\dot{\phi
}(t)^{2}}{
\Lambda^{4}_2}-\frac{6 H(t)\dot{\phi}(t)}{\Lambda^{3}_3}+\frac{90 
H(t)^{2}\dot{\phi}(t)^{2}}{\Lambda^{6}_{3}}\right]
\label{eq2}\\
&&\!\!\!\!\!\!\!\!\!\!\!
\left[M_{pl}^{2}\dot{H}(t)+\frac{\rho_m(t)}{2}+\frac{p_m(t)}{2}
\right]\left[1-\frac{\dot{\phi}(t)^4}{\Lambda^8_2}   
\right]=M_{pl}\, 
\frac{\dot{\phi}(t)^2}{\Lambda^4_2} 
\left[1-24\frac{H(t)\dot{\phi}(t)}{\Lambda_{3}^{3}}\right] \ddot{\phi}(t)\nonumber\\
&& \ \ \ \ \ \ \ \ \ \ \ \ \ \ \ \ \ \ \ \ \ \ \ \ \ \ \ \ \ \ \ \ \ \ \ \ \ \ \ \ \ \ \ 
\ 
\ \ \ \ \ \ \ \,
- \dot{\phi}(t)^2  F\left(\dot{\phi}(t)\right),
\label{eq3}
\end{eqnarray}
while using (\ref{f}) the function $F(X,Z)$ reads as
\begin{equation}
F\left(\dot{\phi}(t)\right)=\frac{1}{2}+2 \frac{\dot{\phi}^2}{\Lambda^4_2} 
-\frac{3H\dot{\phi}}{\Lambda_{3}^{3}}+36\frac{H^{2}\dot{\phi}^{2}}{
\Lambda_{3}^{6}}.
\label{eq4}
\end{equation}
Similarly, the scalar-field equation  (\ref{scalareom}) becomes
\begin{eqnarray}
\frac{1}{a(t)^3}\frac{d}{dt}\left[2a(t)^{3}\dot{\phi}(t)F\left(\dot{\phi}(t)\right) 
\right]&=&-\frac{\dot{ V } (\phi(t)) } {
\dot { \phi } (t) } \label{eq1}.
\end{eqnarray}
Note that we have considered all quantities in the above equations to depend on $t$, 
and $a(t)$, $H(t)$, $\dot{H}(t)$ are given by 
(\ref{matterbounce}),(\ref{Htt}),(\ref{Hdottt}).

As we can see, the second Friedmann equation  (\ref{eq3}) is independent of the 
potential $V(\phi(t))$. Hence, once the matter equation-of-state parameter is given (in 
which case (\ref{mattconserv}) provides $\rho_m(t)$), Eq.  
(\ref{eq3}) can be used to provide a solution for $\phi(t)$ and $\dot{\phi}(t)$.  In 
particular, Eq. (\ref{eq3}) is a simple differential equation for $\dot{\phi}(t)$, namely
\begin{equation}
 \ddot{\phi}(t)=Q(\dot{\phi}(t),t),
\label{eq5}
\end{equation}
that can be easily solved to find $\dot{\phi}(t)$ and hence $\phi(t)$. 
Similarly, the scalar-field equation (\ref{eq1}) is a simple differential 
equation for $V(t)$ of the form  
\begin{equation}
 \dot{V}(t)=P(\dot{\phi}(t),t).
\label{eq6}
\end{equation}
Thus, substituting the solution for $\dot{\phi}(t)$ into  (\ref{eq6}) and integrating we 
can immediately find $V(t)$. In summary, having found the solution for $\phi(t)$ and 
$V(t)$ we can obtain $V(\phi)$ in a parametric form. Hence, this re-constructed potential 
will be the one that generates the bouncing scale factor (\ref{matterbounce}). As 
we described above, we mention that the freedom to have a potential allows to obtain a 
bounce even for simple and not tuned  $G_I$'s, as those chosen in  (\ref{ansGI1}),
which is the motivation of the present paragraph.

\begin{figure}[ht]
\centering
\includegraphics[width=7.4cm,height=5.5cm]{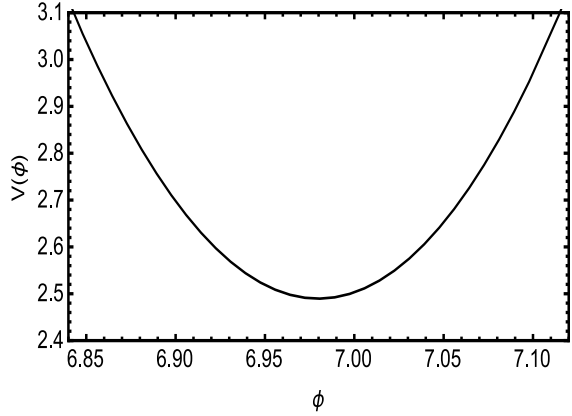}
\caption{\small {\em The reconstructed scalar potential $V(\phi)$ 
that generates the bouncing scale factor (\ref{matterbounce}), in the case where 
$G_{2}=G_{4}=X^2$, $G_3=X$, $G_5=0$. The bouncing parameters have been chosen as  
$ a_b=0.2$, $B=10^{-5}$, while $\Lambda_{2}=0.9$,  $\Lambda_{3}=0.01$, in $M_{pl}$ units. 
}}
\label{potentialscalefactorgiven}
\end{figure}

In general the above procedure cannot be performed analytically, due to the complicated 
forms of the involved equations. Therefore, in order to provide a concrete example, we 
proceed to a numerical application of the above steps. Moreover, since we desire to 
investigate the pure effect of the novel terms of action (\ref{fullaction}), we neglect 
the matter sector. In Figure 
\ref{potentialscalefactorgiven} we present the
potential $V(\phi)$ that is reconstructed from the given bouncing scale-factor form  
(\ref{matterbounce}), according to the above procedure. 

\begin{figure}[ht]
\centering
\includegraphics[width=8cm,height=5cm]{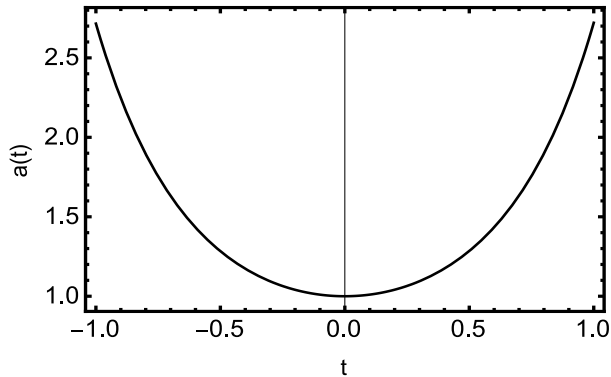}
\caption{\small  {\em The evolution of the scale factor $a(t)$ that is generated by the 
simple parabolic potential (\ref{potential}), in the case where 
$G_{2}=G_{4}=X^2$, $G_3=X$, $G_5=0$. The potential parameters have been chosen as  
$  V_{0}=8.5$, $\phi_{0}=7.0$, while $\Lambda_{2}=0.9$,  $\Lambda_{3}=0.01$, in $M_{pl}$ 
units.
}  }
\label{scalefactorpotentialgiven}
\end{figure}

As we can see from Figure \ref{potentialscalefactorgiven}, in order to 
obtain a bouncing scale factor in the case where $G_{2}=G_{4}=X^2$, $G_3=X$, we need a 
potential with a simple minimum. Hence, we can now reverse the reconstruction procedure 
and consider a potential of the simple form
\begin{equation}
V(\phi)=V_0+(\phi-\phi_{0})^2,
\label{potential}
\end{equation} 
where $V_{0}$ and $\phi_{0}$ are parameters. Inserting this form into Eqs. 
(\ref{eq2})  
and (\ref{eq1}), we obtain a system of two ordinary differential equations for $a(t)$ and 
$\phi(t)$, that can be easily solved numerically. In Figure 
\ref{scalefactorpotentialgiven} we depict the scale factor $a(t)$ that results from the 
given potential (\ref{potential}). Hence, we indeed verify that the simple 
parabolic potential (\ref{potential}) can generate a cosmological bounce.
We  mention that the above procedures can be straightforwardly 
applied in the case where the matter sector is present, i.e describing a matter bounce. 
In particular, one can repeat the above steps, with the inclusion of a  pressureless 
matter, i.e. with  $p_{m}=0$  and $\rho_{m0}=\rho_{m0}/a^{3}$, where $\rho_{0}$ is the 
matter energy density at the time of the bounce.  

\subsubsection{Reconstructing $G_3(X)$}
\label{reconstructionb222}

In this paragraph we chose a priori three out of the four $G_I$'s, and we set the 
potential to be zero or to have a specific given form, and we suitably reconstruct the 
remaining $G_I$ in order to obtain the bouncing solution (\ref{matterbounce}).
Without loss of generality we determine by hand $G_2$, $G_4$ and $G_5$ without any 
particular tuning, i.e we chose
\begin{equation}
G_{2}=G_{4}=X^2;\ \ \   G_5=0,
\label{ansGI1222}
\end{equation}
and furthermore we assume that the potential $V(\phi)$ is determined too. 
Inserting (\ref{matterbounce}) and (\ref{ansGI1222}) into the Friedmann equations 
(\ref{fried1}),(\ref{fried2}) we obtain 
\begin{equation}
3M_{pl}^{2}H(t)^{2}=\rho_m(t)+V(\phi(t))+\dot{\phi}(t)^2\left[\frac12+\frac{3\dot{\phi}
(t)^2}{\Lambda_2^4}-\frac{6H(t)\dot{\phi}(t)}{\Lambda_3^3}G_{3X}(t)+\frac{90H(t)^2\dot{
\phi } (t)^2}{ \Lambda_3^6}
\right]
\label{eq2222}
\end{equation}
\begin{eqnarray}
&&\!\!\!\!\!\!\!\!\!\!\!
\left[M_{pl}^{2}\dot{H}(t)+\frac{\rho_m(t)}{2}+\frac{p_m(t)}{2}
\right]\left[1-\frac{6\dot{\phi}(t)^4}{\Lambda^8_2}   
\right]=M_{pl}\frac{\dot{\phi}(t)^2}{\Lambda_{2}^4}
\left[G_{3X}(t)-24\frac{H(t)\dot{\phi}(t)}{\Lambda_3^3}\right]
\ddot{\phi}
(t)
\nonumber \\
&& \ \ \ \ \ \ \ \ \ \ \ \ \ \ \ \  \ \ \ \ \ \ \ \ \ \ \ \ \ \ \ \ \ \ \ \ \ \ 
\ 
\ \ \ \ \ \ \ \,
-\dot{\phi}(t)^2 
F\left(\dot{\phi}(t)\right),
\label{eq3222}
\end{eqnarray}
while using (\ref{f}) the function $F(X,Z)$ becomes
\begin{equation}
F\left(\dot{\phi}(t)\right)=\frac12+\frac{2\dot{\phi}(t)^2}{\Lambda_2^4}-\frac{3H(t)\dot{
\phi}(t)}{\Lambda_3^3}G_{3X}(t)+\frac{36H(t)^2\dot{\phi}(t)^2}{\Lambda_3^6}.
\label{eq4222}
\end{equation}
Similarly, the scalar-field equation  (\ref{scalareom}) writes as
\begin{eqnarray}
\frac{1}{a(t)^3}\frac{d}{dt}\left[2a(t)^{3}\dot{\phi}(t)F\left(\dot{\phi}(t)\right) 
\right]&=&-\frac{ d  V(\phi)} { d \phi }(t) \label{eq1222}.
\end{eqnarray}
Note that we have considered all quantities in the above equations to depend on $t$, 
and $a(t)$, $H(t)$, $\dot{H}(t)$ are given by 
(\ref{matterbounce}),(\ref{Htt}),(\ref{Hdottt}).

Equations (\ref{eq2222}), (\ref{eq3222}) and (\ref{eq1222}), out of which only two are 
independent, form a system of differential equations for $\dot{\phi}(t)$ and $G_{3X}(t)$. 
Eq. (\ref{eq2222}) can be immediately algebraically solved in terms of $G_{3X}(t)$, and 
thus insertion into (\ref{eq4222}) and then into  (\ref{eq1222}), once the matter 
equation-of-state parameter is given, leads to  a simple second-order differential 
equation for $\phi(t)$, namely
\begin{equation}
 \ddot{\phi}(t)=S(\dot{\phi}(t),\phi(t),t),
\label{eq5222}
\end{equation}
that can be easily solved to find  $\phi(t)$. Then, insertion of  $\phi(t)$ back in 
(\ref{eq2222}) gives $G_{3X}(t)$. Finally, knowing  $\phi(t)$, i.e  $\dot{\phi}(t)$, i.e 
$X(t)$, as well as  $G_{3X}(t)$ allows us to reconstruct  $G_{3X}(X)$ and by integration 
$G_3(X)$. Note that the above procedure is significantly simplified in the case where the 
potential is absent, since then   (\ref{eq5222}) becomes an algebraic equation for 
$\dot{\phi}$, namely 
\begin{equation}
a(t)^3\dot{\phi}(t) 
\left[\frac{1}{4}+\frac{\dot{\phi}(t)^2}{2\Lambda_2^4}-\frac{9H(t)^2\dot{
\phi}(t)^2}{\Lambda_3^6} +\frac{   M_{pl}^{2}H(t)^2 -\rho_m(t)   }{2\dot{
\phi}(t)^2}
\right]=const.,
\label{algbracV0}
\end{equation}
which then leads to an easy determination of $G_{3X}(t)$.

In summary, the above procedure allows us to reconstruct $G_{3}(X)$, which will be the 
one that generates the bouncing scale factor (\ref{matterbounce}). As we described above, 
we mention that fixing the potential or taking it to be zero, leads to a complicated form 
of one of the $G_I$'s (in our example of $G_3$) in order for the bouncing solution to be 
realized. 

Hence, one can now clearly see the difference in the procedure of the present 
paragraph, with that of paragraph \ref{reconstructionb111}. In the present analysis the 
bounce is obtained with a simple or zero potential but with a suitably 
reconstructed, complicated $G_3$, while in paragraph \ref{reconstructionb111} the bounce 
was obtained with simple  $G_I$'s, but with a suitably 
reconstructed, complicated potential. 

\begin{figure}[ht]
\centering
\includegraphics[width=7.6cm,height=5.5cm]{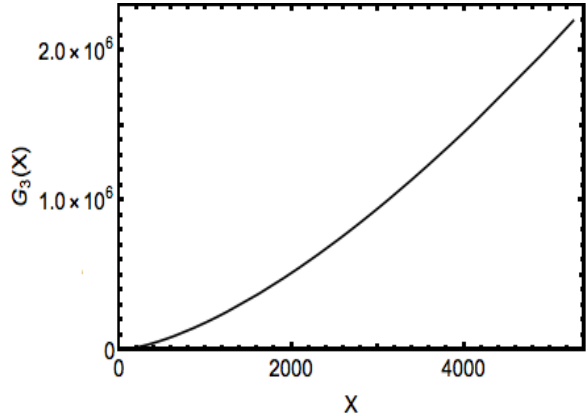}
\caption{\small {\em  The reconstructed galileon function  $G_3(X)$ 
that generates the bouncing scale factor (\ref{matterbounce}), in the case where 
$V(\phi)=0$, and with $G_{2}=G_{4}=X^2$, $G_5=0$. The bouncing parameters have 
been chosen as $ a_b=0.2$, $B=10^{-5}$, while $\Lambda_{2}=0.9$,  $\Lambda_{3}=0.01$, in 
$M_{pl}$ units. }}
\label{G3reconstr}
\end{figure}

In general the above procedure cannot be performed analytically, due to the complicated 
forms of the involved equations. Therefore, in order to provide a concrete example, we 
proceed to a numerical application of the aforementioned steps. Furthermore, since we 
desire to investigate the pure effect of the novel terms of action (\ref{fullaction}), 
we neglect the matter sector. In Fig. \ref{G3reconstr} we present the
function $G_3(X)$ that is reconstructed from the given bouncing scale-factor form  
(\ref{matterbounce}), according to the above procedure, in the case where the scalar 
potential is zero. As we can see, even in the case of zero potential, with simple choices 
for the three  $G_I$'s, a bounce can still be realized if one uses a suitably 
reconstructed, complicated  galileon function  $G_3$. 
 
\subsubsection{Analytical conditions for bouncing solutions}
\label{reconstructionb333}

We close this subsection by investigating analytical bouncing solutions in the case of 
matter absence. In particular, 
substituting \eqref{Z_def1} into the first Friedmann equation \eqref{fried1} we obtain 
the general equation satisfied by the Hubble function, namely
\begin{equation}
aH^3+bH^2+cH+d=0,
\end{equation}
where $a,b,c,d$ are time-dependent coefficients given by
\begin{eqnarray}
a &=& \frac{2\dot{\phi}}{\Lambda_{3}^{9}}\left(\frac{5G_{5X}}{X}+2G_{5XX}\right) \\
b &=& 
-\frac{6\dot{\phi}^{2}}{\Lambda_{3}^{6}}\left(\frac{G_{4}}{X^{2}}-\frac{4G_{4X}}{X}-4G_{
4XX }
\right)-3M_{
pl}^{2}\\
c &=& -\frac{6\dot{\phi}}{\Lambda_{3}^{3}}G_{3X}\Lambda_{2}^{4}X \\
d &=& V+\frac{\Lambda_{2}^{4}X}{2}-G_{2}\Lambda_{2}^{4}.
\label{conditions}
\end{eqnarray}
The general solution of the above cubic equation is
\begin{eqnarray}
&&H 
=-\frac{b}{3a}-\frac{2^{3/2}(3ac-b^{2})}
{3a\left[9abc-2b^3-27a^{2}d+(9abc-2b^3
-27a^ { 2 } d)\sqrt{4(3ac-b^2)^{
3}}\right]^{1/3}} \nonumber \\
 && \ \ \ \ \ \ \, 
+2^{-3/2}\left[9abc-2b^3-27a^{2}d+(9abc-2b^3
-27a^ { 2 } d)\sqrt{4(3ac-b^2)^{
3}}\right]^{1/3}.
 \label{general}
\end{eqnarray}

According to the discussion of this subsection, the general bounce requirements are $H=0$ 
and $\dot{H}>0$ at the bounce point. Hence, using (\ref{general}), the first requirement, 
namely  $H=0$, gives us the conditions
 \begin{eqnarray}
 \label{b1}
b^{2}&=&3ac ; d=0 \\
&&\quad {\rm or} \nonumber  \\
b^{2}&=&3ac ; d=\frac{b^{3}}{18a^{2}} \label{b2},
\end{eqnarray}
which must hold at the bounce moment. On the other hand, the second requirement, namely  
$\dot{H}>0$, using 
\eqref{fried2} leads to the condition
\begin{equation}
\frac{\left(\Lambda_{2}^{4}X+2\Lambda_{2}^{4}XG_{2X}+2M_{pl}\ddot{\phi}XG_{3X}\right)}{
(4XG_ {
4X}-2G_{4}
-1)}>0,
\label{secondcond}
\end{equation} 
around the bouncing point.

Let us make some comments on the conditions \eqref{b1} or \eqref{b2}  and 
(\ref{secondcond}). As we observe, these conditions depend mainly on the functions 
$G_I$'s, however they depend on $V(\phi)$ too, since coefficient $d$ in  \eqref{b1} or 
\eqref{b2} includes  $V(\phi)$, while $\ddot{\phi}$ that appears in (\ref{secondcond}) 
depends on $V(\phi)$ through the Klein-Gordon equation (\ref{scalareom}). Nevertheless, 
in the case where all $G_I$'s are zero there is no potential that can realize 
both conditions, which is expected since in the case of minimally-coupled general 
relativity the null energy condition cannot be violated and thus a bounce cannot be 
realized. On the other hand, in the case where the potential is zero, there are suitable  
$G_I$'s that can satisfy  \eqref{b1} or \eqref{b2}  and (\ref{secondcond}) and induce a 
bounce. From these we deduce that the crucial ingredient of bounce realization is the 
Galileon functions  $G_I$'s and not the potential $V(\phi)$. However, this does not 
forbid one to consider a non-zero potential in order to alleviate some tuning 
from the functions $G_I$'s, or in order to transfer all the required tuning from the 
 $G_I$'s to the suitably reconstructed, complicated  $V(\phi)$.

Let us consider the case where the potential is non-zero and free to be suitably 
reconstructed.
Observing conditions \eqref{b1} and \eqref{b2},  one can easily see that 
the simplest 
model of weakly broken galileon theories possible to generate a bounce must have the 
first three $G_I$ functions non-zero, namely  $G_{2}\neq 0, G_{3}\neq 
0, G_{4}\neq 0$ and $G_{5}=0$, since if $G_5=0$ then the condition $b^{2}=3ac$ cannot be 
satisfied if $G_4=0$. In this simplest model, at the bounce point we have $a=b=0$ and 
thus \eqref{b1},\eqref{b2} imply that at the bounce point:
\begin{eqnarray}
\label{bcond1}
\dot{\phi}^{2}|_{b}&=&\frac{M_{pl}^{2}\Lambda_{3}^{6}}{2\left(\frac{4G_{4X}}{X}+4G_{4XX}
-\frac { G_ { 4 }
}{X^{2}}\right)}
\\
V(\phi)|_{b} &=& G_{2}\Lambda_{2}^{4}-\frac{\Lambda_{2}^{4}X}{2}.
\end{eqnarray}
Additionally, using the solution \eqref{general}, we deduce that before the bouncing 
point ($H<0$) we must have $b>0 $, while after the bouncing point ($H>0$) we must have 
$b<0 $, or equivalently
\begin{eqnarray}
\label{bcondexp}
\dot{\phi}^{2}&<&\frac{M_{pl}^{2}\Lambda_{3}^{6}}{2(\frac{4G_{4X}}{X}+4G_{4XX}-\frac{G_{4}
}{X^{2}})}
 \quad {\rm for\ expansion}\\
\dot{\phi}^{2}&>&\frac{M_{pl}^{2}\Lambda_{3}^{6}}{2(\frac{4G_{4X}}{X}+4G_{4XX}-\frac{G_{4}
}{X^{2}})}
 \quad {\rm for\ contraction}.
 \label{bcondcon}
\end{eqnarray}

Let us apply these in the model  (\ref{ansGI1}), which indeed belongs to the subclass of 
simplest models considered here. In this case (\ref{bcond1}) becomes: 
\begin{equation}
\dot{\phi}^{2}|_{b}=\frac{M_{pl}^{2}\Lambda_{3}^{6}}{30},
\label{atbounce}
\end{equation}
while (\ref{bcondexp}),(\ref{bcondcon}) become respectively
\begin{eqnarray}
\dot{\phi}^{2}&<&\frac{M_{pl}^{2}\Lambda_{3}^{6}}{30 }\quad {\rm for\ 
expansion}\\
\dot{\phi}^{2}&>&\frac{M_{pl}^{2}\Lambda_{3}^{6}}{30 } \quad {\rm for\ 
contraction},
\label{atexpcont}
\end{eqnarray}
and finally (\ref{secondcond}) reads as
\begin{equation}
\frac{\dot{\phi}^{2}\Lambda_{2}^8+4\Lambda_{2}^{4}\dot{\phi}^{4}+2M_{pl}\ddot{\phi
}\dot{\phi}^{2}\Lambda_{2}^{4}}{6\dot{\phi}^{4}-\Lambda_{2}^8}>0.
\label{atexpcont222}
\end{equation}
 The most general form of $\dot{\phi}$ which satisfies  \eqref{atbounce}, 
\eqref{atexpcont} and \eqref{atexpcont222} is
\begin{equation}
\dot{\phi}=\alpha t^{\gamma}+\beta,
\label{e6}
\end{equation}
where  $\gamma=1,3,5,..$, $\beta=M_{pl}\Lambda_{3}^3/\sqrt{30}$ and $\alpha$ a negative 
constant. In order to give a simple example let us choose $\gamma=1$. Integrating the 
above expression we obtain 
\begin{equation}
\phi(t)={\alpha t^{2}\over 2}+\beta t+\delta,
\label{eq7}
\end{equation}
with $\delta$ an integration constant. Substituting   \eqref{eq7} into the 
first Friedmann equation  \eqref{eq2} we acquire 
  \begin{equation}
V(t)=
3H(t)^2M^{2}_{pl}-(t\alpha+\beta)^{2}\left[ 
\frac{1}{2}+\frac{3(t\alpha+\beta)^{2}}{\Lambda^{4}_2}  +\frac{6 
H(t)(t\alpha+\beta)^2(15H(t)-\Lambda^{3}_3)}{\Lambda^6_3}\right],
\label{eq8}
\end{equation}
Additionally, the 
second Friedmann equation \eqref{eq3} can provide the solution for $H(t)$. Hence, one can 
eliminate time,  obtaining a general form of 
the potential $V(\phi)$ that generates a bouncing evolution.

Let us make a comment here on the role of the parameters $\Lambda_2$ and  $\Lambda_3$ 
that characterize the weakly broken galileon invariance. As we mentioned above 
$\Lambda_{3}$ marks the scale suppressing the invariant galileon interactions, while the 
parameter $\Lambda_{2}$,  with $\Lambda_{3}\ll \Lambda_{2}$,
marks the  scale suppressing the quantum-mechanically generated single-derivative
operators \cite{Pirtskhalava:2015nla}, and thus in the limit where both 
$\Lambda_{2}$,$\Lambda_{3}$ go to $M_{pl}$ weakly broken galileon invariance 
disappears, and the above theories become the usual covariant galileon ones. Hence, we 
can clearly see that the freedom to set the values of $\Lambda_2$, $\Lambda_3$ 
semi-independently makes the theories at hand different than usual covariant 
galileon ones, and the corresponding cosmology richer. In particular, one can see that 
the above bouncing requirements are much more difficult to be fulfilled  
in the case where 
$\Lambda_{2}$,$\Lambda_{3}$ are set to $M_{pl}$, and similarly the specific 
numerical examples of the previous paragraphs would be harder to be provided. We mention 
however that the comparison of the theories with  weakly broken galileon invariance is 
made in relation to usual covariant galileon theories, and not with the general 
Horndeski theory, since as it was discussed in \cite{Pirtskhalava:2015nla} the theories 
at hand fall within the class of general Horndeski construction.

We close this subsection by discussing on an issue that is present in principle in almost 
every bouncing scenario, namely the anisotropy issue. In particular, the bounce 
realization is in principle unstable against  anisotropic stress, the so-called  BKL 
instability \cite{Belinsky:1970ew}, since   the effective energy density in anisotropies 
$\rho_{anis}$ evolves proportionally to  $a(t)^{-6}$ and therefore in a contracting 
universe it increases   faster than the matter and radiation energy densities, and thus 
the bounce should be realized in a non-isotropic and non-homogeneous 
spacetime.

However, although anisotropies grow faster than the background evolution, they will not 
dominate quickly, since this is related to the initial conditions of the anisotropy 
generation. Specifically, if the anisotropies arise from the backreaction of cosmological 
perturbations, which is of quantum origin, then the moment that the anisotropies will 
dominate over the background depends on the energy scale of the universe during matter 
contraction, which is typically at quite high energy scales. Hence, one can reliably 
consider an FRW background evolution up to the bounce phase.

Apart from this, there are many mechanisms that can be additionally introduced in order 
to ensure that even at high energy scales the anisotropies will not dominate and make 
the universe depart from its FRW evolution. A well-studied case is to realize an 
ekpyrotic scenario, through the introduction of a negative exponential potential
(see 
\cite{Cai:2012va,Qiu:2013eoa,Cai:2013vm,Cai:2013kja} for the details of such a 
construction and 
\cite{Cai:2014xxa,Cai:2014bea} for its observational confrontations), and thus  it can be 
straightforwardly introduced in the present galileon scenario too (though one should be 
careful not to destroy the background bouncing solution, i.e. he should follow the 
procedure of subsection \ref{reconstructionb222}).

Having these in mind, in the following section, where we extract the observables of the 
bounce phase, we perform our analysis in the FRW geometry, without the need to examine a 
non-isotropic background evolution.

\subsection{Reconstruction of cyclic evolution}

Let us now extend the above analysis constructing a sequence of bounces and turnarounds, 
i.e examining the  realization of cyclic evolution.
Without loss of generality we consider an oscillating scale factor of the form  
 \begin{equation}
 \label{cyclicscaefactor}
 a(t)=A \sin(wt)+a_{c},
 \end{equation}
  where  $a_{c}-A>0$ is the scale factor value at the bounce, with  $A+a_c$ the scale 
factor value at the turnaround. In this case we apply the reconstruction procedure of the 
previous subsection, namely relations (\ref{eq2})-(\ref{eq6}),  in order to extract the 
solutions 
for $\phi(t)$ and $V(t)$, and thus obtain the re-constructed potential $V(\phi)$. Hence, 
this re-constructed potential will be the one that generates the cyclic scale factor 
(\ref{cyclicscaefactor}). Note that the matter sector has to been considered in this case,
hence we can assume it to be dust, namely with $p_m=0$ and with $\rho _{m}=\rho _{mb} 
(a_{c}-A)^{3}/a^{3}$, with $\rho _{mb}$ the value at the bouncing point.

\begin{figure}[ht]
\centering
\includegraphics[width=7.4cm,height=5.5cm]{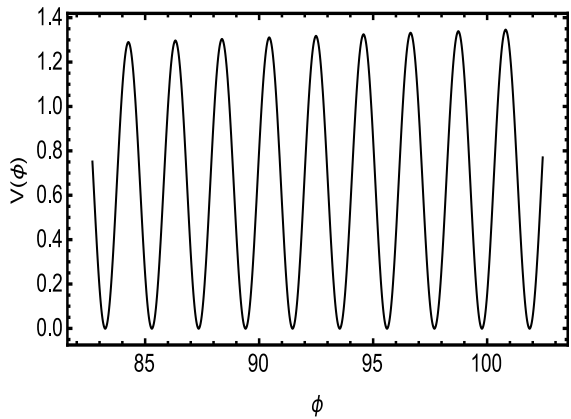}
\caption{\small {\em The reconstructed scalar potential $V(\phi)$ 
that generates the cyclic scale factor 
(\ref{cyclicscaefactor}), in the case where 
$G_{2}=G_{4}=X^2$, $G_3=X$, $G_5=0$. The model parameters have been chosen as  
$ a_c=0.01$, $A=10^{-4}$, $w=15$, $\rho_{mb}=0.01$, while $\Lambda_{2}=0.9$,  
$\Lambda_{3}=0.01$, in $M_{pl}$ units. }}
\label{fig:potential1}
\end{figure}

In order to provide a concrete example we proceed to a numerical application of the 
above steps. In Figure \ref{fig:potential1} we present the
potential $V(\phi)$ that is reconstructed from the given  cyclic scale-factor form  
(\ref{cyclicscaefactor}), according to the above procedure, in the case where 
$G_{2}=G_{4}=X^2$, $G_3=X$. 

\begin{figure}[ht]
\centering
\includegraphics[width=8cm,height=5cm]{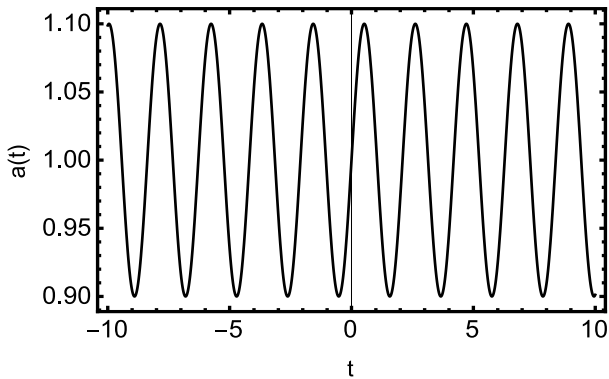}
\caption{\small  {\em The evolution of the scale factor $a(t)$ that is generated by the 
simple oscillatory potential (\ref{potentialcyclic}), in the case where 
$G_{2}=G_{4}=X^2$, $G_3=X$, $G_5=0$. The potential parameters have been chosen as  
$V_{1}=1$, $V_{2}=0.1$, and $w_{V}=3$, the matter energy density at the bounce as  
$\rho_{mb}=0.01$, while $\Lambda_{2}=0.9$,  
$\Lambda_{3}=0.01$, in $M_{pl}$ 
units.}}
\label{fig:scalefactor1}
\end{figure}

As we can see from Figure \ref{fig:potential1}, in order to 
obtain a cyclic scale factor in the case where $G_{2}=G_{4}=X^2$, $G_3=X$, we need a 
potential with an oscillatory form. Hence, we can now reverse the reconstruction 
procedure and consider a potential of the simple form
\begin{equation}
\label{potentialcyclic}
V(t)=V_{1}\sin(w_{V}t)+V_{2},
\end{equation}
where $V_1$, $V_2$ and $w_V$ are parameters. As in the bounce reconstruction, inserting 
this form into Eqs. (\ref{eq2}) and (\ref{eq1}), we obtain a system of two ordinary 
differential equations for $a(t)$ and $\phi(t)$, that can be easily solved numerically. 
In 
Figure 
\ref{fig:scalefactor1} we depict the scale factor $a(t)$ that results from the 
given potential (\ref{potentialcyclic}). Thus, we indeed verify that the simple 
oscillatory potential (\ref{potentialcyclic}) can generate a cyclic universe.

Finally, we close this subsection by investigating some analytical cyclic solutions. A 
possible form of the scalar field $\phi$ which is able to satisfy the 
conditions at and around the bounce given by \eqref{atbounce} and \eqref{atexpcont}, and 
is also oscillatory in nature, reads as
\begin{equation}
\phi(t)=p \frac{\sin(wt)}{w}+\frac{s t^{2}}{2}+t l +c_0,
\label{eqq8}
\end{equation} 
where $p$, $w$, $s<0$ and $l$ are parameters and $c_0$ an integration constant. 
  Substituting  \eqref{eqq8} either into  
\eqref{eq1} or into  \eqref{eq2}, we 
obtain
\begin{eqnarray}
&&\!\!\!\!\!\!\!\!\!\!\!\!\!\!\!\!\!\! 
V(t)=
3H(t)^2M^{2}_{pl}-[1+st+p\cos(wt)]^{2}\left\{
 \frac{6 
H(t)[1+st+p\cos(wt)]^2[15H(t)-\Lambda^{3}_3]}{\Lambda^6_3}
 \right. \nonumber \\ && 
\ \ \ \ \ \ \ \ \ \ \ \ \ \ \ \ \ \ \ \ \ \ \ \,
 ~~~~~~~~~~~~~~~~~~~~~~~ \left.
 +\frac{1}{2}+\frac{3[1+st+p\cos(wt)]^{2}}{\Lambda^{4}_2}
 \right\},
\label{eqq9}
\end{eqnarray}
while \eqref{eq3} can give the solution for $H(t)$. Hence, one can 
eliminate time,  obtaining 
the potential $V(\phi)$ that generates a cyclic evolution.

In this subsection we showed that at the background level the theories with weakly 
broken galileon symmetry can give rise to cyclic cosmology. However, we stress that such 
a possibility has mainly a theoretical interest in order to reveal the capabilities of 
the scenario, since these cyclic scenarios will suffer from the problems of every cyclic 
evolution concerning perturbation-related observables, such as the spectral index. In 
particular, as it was shown in \cite{Brandenberger:2009ic},  in every cyclic cosmological 
model at each cycle fluctuations grow on super-Hubble scales during the contracting phase, 
and this induces a jump in the curvature power-spectrum spectral index $n_s$ by $\delta 
n_s=-2$, and hence these models lose predictability. Thus, in order to consider these 
models as realistic, one should extend them and incorporate mechanisms that could 
alleviate this problem, for instance through a long dark-energy period before the 
turnaround of each cycle, as it was done in \cite{Piao:2010cf}.

\section{Cosmological Perturbations in the bounce phase}
\label{perturbations}

In   subsection \ref{reconstructionb} we investigated the bounce realization in the 
framework of weakly broken galileon theories at the background level. In this section we 
proceed to the investigation of perturbations. Such a study is necessary in every 
bouncing scenario, since, similarly to inflationary cosmology, they will be related to 
observations. 

The usual process for generating the primordial power spectrum in inflationary cosmology
requires that cosmological fluctuations initially emerge inside the Hubble radius, then 
they exit it in the primordial epoch, and finally they re-enter at late times 
\cite{Starobinsky:1979ty}. In bouncing cosmology however, the quantum
fluctuations around the initial vacuum state are generated well in advance of the 
bouncing phase, and as contraction continues they exit the Hubble radius, since 
the wavelengths of the primordial fluctuations decrease slower than the Hubble radius. 
Definitely, when the universe passes through the bounce point the  background evolution 
could affect the perturbations scale-dependence mainly in the UV, however the IR regime, 
which is responsible for  the observable primordial perturbations related to the 
large-scale structure, will remain almost unaffected since at this regime the 
gravitational 
modification effects are very restricted 
\cite{Battarra:2014tga,Quintin:2015rta,Koehn:2015vvy}. Hence, one can study 
the primordial power-spectrum formation within standard cosmological perturbation 
theory.

Let us start by analyzing the perturbations in the framework of weakly broken galileon 
theories \cite{Pirtskhalava:2015nla}. As usual, we consider that at linear order  scalar 
and tensor perturbations decouple and evolve independently, and moreover note that for 
the 
present class of theories, which form a subclass of Horndeski theory, the equation of 
motion for the scalar field is still of second order. One novel feature of the present 
scenario is that apart from the usual symmetries present in FRW geometry, we 
additionally have the weakly broken galileon invariance. Hence, in the following we will 
see its effect on the perturbations.

We follow the usual Arnowitt-Deser-Misner (ADM) formalism, in which the metric is 
decomposed as 
\begin{equation}
ds^2=-N^2 dt^2+h_{ij}(N^i dt+dx^i)(N^j dt+dx^j),
\end{equation}
where  $N=1/\sqrt{-g^{00}}$ is the lapse and $N^i$ the shift functions, while $h_{ij}$ is 
the 3$D$ metric on constant time hypersurfaces. In order to study the perturbations, we 
need to expand the action up to quadratic order in metric fluctuations. The intrinsic 
curvature of equal-time hypersurfaces, i.e. $^{(3)}R$, is at least linear in 
perturbations, while the 
extrinsic curvature of equal-time 
hypersurfaces, defined as
\begin{equation}
\label{Kij}
K_{ij} = \frac{1}{2N}(\dot{h}_{ij} - \nabla_iN_j-\nabla_jN_i)
\end{equation}
where the covariant derivative $\nabla_i$ are taken with respect to  $h_{ij}$,
must be perturbed around the flat FRW background. Hence, we consider
\begin{eqnarray}
N &=&1+\delta N ,\nonumber \\
K_{ij} &=& H h_{ij}+\delta K_{ij} .
\end{eqnarray}
The perturbed action   then reads
\cite{Pirtskhalava:2015nla,Piazza:2013coa,Gleyzes:2013ooa}
\begin{equation}
\label{pertaction}
\begin{split}
S =   \int & \ d^4x \sqrt{ \gamma }\, N \, \bigg\{\frac{M_{pl}^2 }{2} f(t) 
\Big[{}^{(3)}\!R+K^{ij}K_{
ij}-K^2\Big] - 2 \dot f (t) \frac{K}{N}+\frac{c(t)}{N^2} - \Lambda(t)
 \\
& +\frac{M^4(t)}{2}  \delta N^2 -\hat{M}^3(t)\delta K \delta N -\frac{\bar{M}^2(t)}{2} 
\big( \delta 
K^2 - \delta K^{ij} \delta K_{ij} \big)+\frac{\tilde m^2 (t)}{2} \, {}^{(3)}\!R  \, 
\delta 
N  \\&-\frac{\bar M'{}^2(t)}{2} \big( \delta K^2 + \delta K^{ij} \delta K_{ij} \big) 
+m_1(t) ^{(3)}\!R\delta K+ \dots  \bigg\}~.
\end{split}
\end{equation}
The terms in the first line corresponds to zeroth and first order perturbations, whereas 
the rest of the terms are second order in perturbations (we neglect terms giving rise to 
higher order perturbations). The time dependent coefficient $f(t)$ can be always removed 
through a conformal transformation and thus we set it to 1. The quantities $M^4(t), 
\hat{M}^3(t), \bar{M}^2(t), \dots$, are the various effective field theory coefficients 
whose explicit forms will be fixed using the Horndeski Lagrangian  
\cite{Pirtskhalava:2015nla}. As it was shown in  \cite{Pirtskhalava:2015nla}, one finds 
that $\bar M^2=\tilde 
m^2$, since only the combination $- \delta K^2 + \delta K^{ij} \delta 
K_{ij} +  \, {}^{(3) }\!R  \, \delta N $ appears in the action, which being a redundant 
operator can in turn be omitted by redefining the metric. Therefore, the only non-zero 
effective field theory  coefficients  are 
$M^4(t)$ and $\hat{M}^3(t)$.

In order to extract the equations for scalar and tensor perturbations, we work in the 
unitary gauge, which fixes the time and spatial reparametrization. In this gauge the 
metric and scalar field perturbations are given by \cite{Maldacena:2002vr}
\begin{eqnarray}
\delta \phi=0; \quad h_{ij}=a^2 e^{2\zeta}\delta_{ij},
\end{eqnarray}
where
$\zeta$ parametrizes the scalar fluctuations. In the following subsections we study 
scalar and tensor perturbations separately.

\subsection{Scalar Perturbations}

Working in the unitary gauge, setting all effective field theory coefficients (apart 
from $M^4(t)$ and $\hat{M}^3(t)$) to zero, and using the Hamiltonian and momentum 
constrain equations, one obtains the following quadratic action for the scalar 
perturbations $\zeta$ \cite{Pirtskhalava:2015nla}
\begin{eqnarray}
S_\zeta = \int d^4 x~a^3 A(t) M_{pl}^{2}\bigg [ \dot \zeta^2- 
c_{s}^{2}\frac{(\nabla\zeta)^2}{
a^2}     \bigg ]~,
\end{eqnarray}
where
\begin{eqnarray}
\label{Aexpress}
A &=& \frac{M_{pl}^2 \left(3 \hat M^6+2 M_{pl}^2 
M^4-4M_{pl}^4 \dot H\right)}{\left(\hat{M}^3-2M_{pl}^2 
H\right)^2}, \\
c_{s}^{2}&=&\frac{\left(2 M_{pl}^2 H\hat{M}^3 - \hat{M}^6+2 M_{pl}^2 
\partial_{t} \hat{
M}^3-4M_{pl}^4 \dot H\right)}{\left(3 \hat{M}^6+2 M_{pl}^2 M^4-4M_{pl}^4 \dot H\right)}.
\label{cs2express}
\end{eqnarray} 
For the explicit 
expressions of the  effective field theory    coefficients in 
terms of $G_{I}'s,\ X$ and $\phi$ in the general case, one may refer to 
\cite{Gleyzes:2013ooa}. For the purpose of this work it is adequate to use the 
approximate expressions of the two remaining non-zero effective field theory 
coefficients, namely $M^4$ and $\hat{M}^3$, at cosmological backgrounds, which read as  
\cite{Pirtskhalava:2015nla}
\begin{eqnarray}
\label{approxexpressEFT}
 M^4\sim \hat{M}^3 H\sim  M_{pl}^2 H^2. 
\end{eqnarray}
Following the analysis of the previous section, we again consider the ansatzes 
(\ref{ansGI1}), namely 
$G_{2}=G_{4}=X^2$, $ G_3=X$, $ G_5=0$. Nevertheless, even 
in this simple case, whether  $A$ and 
$c_{s}^{2}$, which have a time-dependence, remain positive or not depends on the 
background solution, as can be clearly seen from (\ref{Aexpress}) and (\ref{cs2express}).

\begin{figure}[ht]
\centering
\includegraphics[width=7.6cm,height=5.5cm]{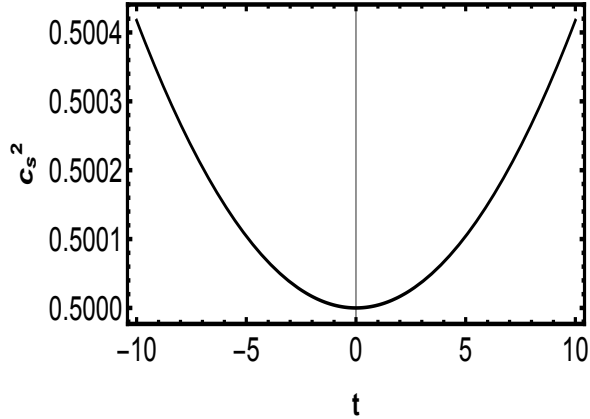}
\caption{\small {\em The evolution of the sound speed square, for the bouncing solution 
of subsection \ref{reconstructionb111}, i.e for the bouncing scale factor 
(\ref{matterbounce}), in the case where $G_{2}=G_{4}=X^2$, $G_3=X$, $G_5=0$, and with   
$ a_b=0.2$, $B=10^{-5}$, $\Lambda_{2}=0.9$,  $\Lambda_{3}=0.01$, in $M_{pl}$ units. 
}}
\label{sound}
\end{figure}

Since $c_{s}^{2}$ in general depends on the background solution, one should  
explicitly check its positivity in any specific example. Concerning for instance the 
specific example of subsection \ref{reconstructionb111}, we insert the ansatz 
for the background bouncing scale factor into  (\ref{Aexpress}) and (\ref{cs2express}) 
and in Fig. 
\ref{sound} we depict the evolution of the sound speed square around the bounce phase. 
Indeed, for this specific example, $c_{s}^{2}$ (and $A$) is positive, and these features 
act in favor of stability.

Proceeding forward, and in order to provide a well-defined perturbation quantization, we 
perform the usual Fourier transformation and introduce the canonical variable 
\begin{equation}
\sigma_{k}=z\zeta_{k}; \quad z=a\sqrt{A}.
\label{standardvariables}
\end{equation} 
Thus, the equation of motion is given by
\begin{equation}
\sigma''_{k}+\left(c_{s}^{2}k^{2}-\frac{z''}{z}\right)\sigma_{k}=0,
\label{scalarmode}
\end{equation}
where primes represent derivatives with respect to conformal time $\eta=\int a^{-1}(t) 
dt$ 
\cite{Qiu:2011cy}. Defining 
\begin{equation}
M^{2}(\eta)=\frac{A'^{2}}{4A^2}-\frac{A''}{2A}-\frac{3HA'}{2A}-\frac{a''}{a},
\end{equation}
we can rewrite the above equation as
\begin{equation}
\sigma''_{k}+\left[c_{s}(\eta)^{2}k^{2}+M^{2}(\eta)\right]\sigma_{k}=0.
\label{scalarmode1}
\end{equation}
In summary, the above equation corresponds effectively to a massive scalar field, whose 
mass and sound speed square are time-dependent, and thus the solution will depend on the 
specific background evolution one imposes.

Let us now apply the obtained background bouncing solutions of the previous section in 
the above equation. As usual, we focus on the contracting phase far away from the 
bounce point, since this is the phase where the scale-invariant power spectra for 
curvature and tensor modes are obtained.
In particular,  for the contracting phase described by  
\eqref{matterbounce}, and far from the bouncing point, where the scale factor evolves as
\begin{equation}
a(t)\approx t^{2/3}\approx \eta^{2},
\end{equation}
 we obtain that 
\begin{eqnarray}
\label{approxexpressEFT}
&&A\simeq M_{pl}^2,\\
&&c_{s}^{2}\simeq1.
\end{eqnarray}
Hence, equation (\ref{scalarmode1}) reduces to 
\begin{equation}
\sigma''_{k}+\left[k^{2}-\frac{2}{\eta^{2}}\right]\sigma_{k}\simeq0.
\label{scalarmode1b}
\end{equation}
At early stages the $k^{2}$-term dominates and hence the gravitational effects can 
be neglected. Therefore, since the scalar fluctuations effectively correspond to a free 
scalar propagating in a flat spacetime, we can consider that the initial 
condition acquires the form of the Bunch-Davies vacuum \cite{Allen:1985ux}:
\begin{equation*}
\sigma_{k}\simeq \frac{e^{-ik\eta }}{\sqrt{2k}}~.
\end{equation*}
Using these vacuum initial conditions we can solve the
perturbation equation (\ref{scalarmode1b}), acquiring 
\begin{equation}
\sigma_{k}=\frac{e^{-ik\eta }}{\sqrt{2k}}\left(1-\frac{i}{k\eta
}\right)~.
\end{equation}
Hence, we deduce that due to the gravitationally-induced term in (\ref{scalarmode1b}), 
after exiting the Hubble radius the quantum fluctuations
could become classical perturbations. Furthermore, the amplitude of the scalar 
perturbations will keep increasing until the moment $t_{bp}$ in which the universe enters 
the bounce phase.

From the definition of the power spectrum we obtain that $\zeta\sim k^{3/2}|\sigma_k|$ is 
scale-invariant in the present scenario. Additionally, the explicit calculation leads to 
a primordial power spectrum of the form
\begin{equation}
P_{\zeta} \equiv \frac{k^3}{2\pi^2}\left|\frac{\sigma_k}{z}\right|^2 \approx
\frac{H_{bp}^2}{48\pi^2M_{Pl}^2},
\end{equation}
where $H_{bp}=\sqrt{{B}/{9}}$ is the absolute value of the
Hubble parameter at $t_{bp}$, i.e.  when the bounce phase starts.

We close this subsection by mentioning that although we performed the perturbation 
analysis in a general way, in order to obtain the power spectrum we focused as usual in 
the contracting phase far away from the bounce, since this is where the 
scale-invariant power spectra for curvature and tensor modes are obtained. Definitely, in 
this regime the action becomes fully canonical, and that is why the analysis and the 
obtained power spectrum coincides with the standard results 
\cite{Novello:2008ra,Finelli:2001sr}. Nevertheless, the general analysis is both 
necessary and interesting if one desires to address the evolution of perturbations 
through the bounce phase. For completeness, we accommodate this issue following 
\cite{Battarra:2014tga,Quintin:2015rta,Koehn:2015vvy}.

If one desires to investigate the processing of perturbations through the bounce phase, 
instead of the variable (\ref{standardvariables}) it proves more convenient to use the   
(gauge-invariant) co-moving curvature perturbation $ \mathcal{R}$ 
\cite{Battarra:2014tga}. In this case, instead of (\ref{scalarmode}) one can write the 
equation of motion for perturbations as
\begin{align}
\label{EOMR}
 \mathcal{R}_k''+2\frac{y'}{y}\mathcal{R}_k'+c_s^2k^2\mathcal{R}_k& = 0,
\end{align}
where 
\begin{eqnarray}
\label{y}
y^2&=& a^2\frac{M_{pl}^2 \left(3 \hat M^6+2 M_{pl}^2 
M^4-4M_{pl}^4 \dot H\right)}{\left(\hat{M}^3-2M_{pl}^2 
H\right)^2}\\
c_{s}^{2}&=&\frac{2 M_{pl}^2 H\hat{M}^3 - \hat{M}^6+2 M_{pl}^2 
\partial_{t} \hat{M}^3-4M_{pl}^4 \dot H}{3 \hat{M}^6+2 M_{pl}^2 M^4-4M_{pl}^4 
\dot H}.
\label{cs2express}
\end{eqnarray}

As it has been shown in \cite{Quintin:2015rta}, for the modes which are of interest 
today, i.e. in the regime $k\ll \mathcal{H}$, the relevant equation is
\begin{equation}
 \frac{d\zeta'}{d\eta}+\frac{(y^2)'}{y^2}\zeta'=0~, 
  \label{EOMy}
\end{equation}
where  $\zeta$ is the perturbation variable, equal to $\mathcal{R}$ in the small-$k$ 
limit, which is the limit we are interested in. One solution of 
(\ref{EOMy}) is  
\begin{equation}
 \zeta'(\eta)=\zeta'(\eta_i)\frac{y^2(\eta_i)}{y^2(\eta)}~,
 \label{solyetaprime}
\end{equation}
with $\eta_i$ the initial time where the initial conditions
are set. Therefore, we deduce that the evolution of perturbations depends on the 
evolution of $y^2$, given by \eqref{y}. Following \cite{Quintin:2015rta}, 
we will examine the behaviour of $y^2$ at three different regimes.

\begin{itemize}
\item  Regime $2|H(t)|\gg |\hat{M}^3|$.

In the regime $2|H(t)|\gg |\hat{M}^3|$ relation  \eqref{y}  in $M_{pl}$ units is 
simplified as
\begin{equation}
y^2=a^2\frac{\left(3 \hat M^6+2M^4-4\dot H\right)}{H^2},
\end{equation}
under the constraint $2M^4+3\hat{M}^6\gg 4\dot H$.
Inserting this expression into \eqref{solyetaprime} we can obtain the evolution of 
perturbations through the bounce, as long as we insert the evolution for the model 
parameters given in \cite{Pirtskhalava:2015nla}, namely (\ref{approxexpressEFT}), for 
our particular background solution of subsections \ref{reconstructionb111} and
\ref{reconstructionb222}.

 \item  Regime $2|H(t)|\ll|\hat{M}^3|$.
 
In the regime $2|H(t)|\ll|\hat{M}^3|$ relation  \eqref{y}  in $M_{pl}$ units is 
simplified as
\begin{equation}
y^2=a^2\frac{\left(3 \hat M^6+2M^4-4\dot H\right)}{\hat{M}^6},
\end{equation}
 (alternatively we could use the parametrization of \cite{Quintin:2015rta} and rewrite the 
Hubble parameter as $H=\alpha \Delta t_B$, where $\Delta t_B$ denotes the bounce 
duration). Similarly to the previous regime, we can obtain the evolution of 
perturbations through the bounce by inserting (\ref{approxexpressEFT}) for the particular 
background solution of subsections \ref{reconstructionb111} and
\ref{reconstructionb222} into \eqref{y} and then into \eqref{solyetaprime}.

 \item  Regime $2|H(t)|\approx|\hat{M}^3$.
 
In this regime, in $M_{pl}$ units, the denominator of \eqref{y} goes to zero leading to 
$y^2\rightarrow \infty$. Thus, after a time interval $t_f$,  the perturbations become 
constant on superhorizon scales. As it has been discussed in detail in  
\cite{Quintin:2015rta, Battarra:2014tga}, the equation of motion seem to become singular 
in this regime, however this feature is an artifact of the Newtonian gauge and is removed 
applying the harmonic gauge. The bounce phase ends after the time $t_f$. Finally, the 
evolution of  perturbations through the bounce is numerically obtained by inserting 
(\ref{approxexpressEFT}) for the particular 
background solution of subsections \ref{reconstructionb111}  and \ref{reconstructionb222} 
into \eqref{y} and then into \eqref{solyetaprime}.

\end{itemize}

\subsection{Tensor Perturbations}

Let us now proceed to the investigation of tensor perturbations following  
\cite{Gleyzes:2013ooa}. As usual, we can neglect the scalar perturbations in 
\eqref{pertaction}. 
Working  in unitary gauge the tensor perturbations read as
\begin{equation}
h_{ij} = a^2(t) e^{2 \zeta} \hat h_{ij} \; , \qquad \det \hat h =1\;, \qquad \hat h_{ij} 
= 
\delta_{
ij} + \gamma_{ij} + \frac12 \gamma_{ik} \gamma_{kj} \;,
\end{equation}
where $\gamma_{ij}$, which parametrizes the tensor perturbation, is assumed to be 
traceless and divergence-free, namely $\gamma_{ii}=0 = \partial_i 
\gamma_{ij}$.  

Using the additional weakly broken galileon symmetry and setting all effective field 
theory coefficients, apart from $M^4(t), \hat{M}^3(t)$, to zero, we acquire the second 
order action for tensor perturbations as
\begin{equation}
S_{\gamma}^{(2)} =\int d^4 x \, a^3 \frac{M_{pl}^2}{8} \left[\dot{\gamma}_{ij}^2 
-\frac1{a^2}(\partial_k \gamma_{ij})^2 \right]
\;.
\end{equation}
Fourier transforming the above equation and working with the canonically normalized 
variable 
$v_{k}=M_{pl}\gamma_{k}/2$, we obtain the equation of motion as
\begin{equation}
v''_{k}+\left(k^{2}-\frac{a''}{a}\right)v_{k}=0.
\label{tensormode}
\end{equation}

Let us now apply the obtained background bouncing solutions of the previous section in 
the above equation. In particular,  for the contracting phase described by  
\eqref{matterbounce}, where the scale factor evolves as
$a(t)\approx t^{2/3}\approx \eta^{2}$,
equation \eqref{tensormode} reduces to
\begin{equation}
v''_{k}+\left(k^{2}-\frac{2}{\eta^{2}}\right)v_{k}=0,
\label{tensormode2}
\end{equation}
whose exact solution is given by
\begin{equation}
v_{k}=\frac{e^{-ik\eta}}{\sqrt{2k\eta}}\left(1-\frac{i}{k\eta}\right).
\end{equation}
 Hence, the primordial power spectrum of tensor fluctuations is also scale-invariant, 
however its magnitude is
 \begin{equation}
P_{T} \equiv \frac{k^3}{2\pi^2}\left|\frac{\sigma_k}{z}\right|^2 \approx
\frac{H_{bp}^2}{48\pi^2M_{Pl}^2},
\end{equation}
which is of the same order of the scalar perturbation. Hence, we deduce that the 
bouncing scenario at hand suffers from the usual problem of all matter-like bounce 
models, namely that the tensor-to-scalar ratio $r\equiv
P_T/P_{\zeta}$ remains of the order one (the scalar power spectrum is not 
additionally amplified as in inflationary realization). 
This high value is in significant 
disagreement with the observed behavior, which according to Planck  probe 
\cite{Planck:2015xua} 
suggests that 
$r< 0.11\, (95\%\,\mathrm{CL})$, while the combined analysis of the BICEP2 and Keck Array 
data with the Planck data requires $r< 0.07\, (95\%\,\mathrm{CL})$ \cite{Array:2015xqh}.  

Note that the above disagreement with the data may be a consequence of a no-go 
theorem that shows that probably all matter-like bounce models would suffer from 
such difficulties in matching observations \cite{Quintin:2015rta}.

In order to accommodate with current observations, and as it is usual in bouncing 
scenarios, we must introduce a mechanism that can magnify the amplitude of 
scalar perturbations, and thus reduce the tensor-to-scalar ratio. For instance one can 
consider an additional light scalar
field, as in the bounce curvaton-bounce \cite{Cai:2011zx}, which can evade the 
aforementioned no-go theorem  and
enhance isocurvature fluctuations, and then give rise to a
scale-invariant spectrum for the adiabatic fluctuations due to kinetic amplification. In 
particular, introducing a massless scalar $\chi$ and considering it to couple to 
the galileon field $\phi$ as $g^2\phi^2\chi^2$, one can follow the procedure of   
\cite{Cai:2011zx} and deduce that the tensor-to-scalar ratio can be reduced to values 
$r\simeq10^{-3}$.

Before closing this section, let us make a comment on the stability of the above 
bouncing solutions. In particular, there is a discussion in the literature whether there 
exists a no-go theorem that forbids stable non-singular cosmologies in 
Horndeski theory, as it was claimed in \cite{Kobayashi:2016xpl,Akama:2017jsa}, which 
could be evaded only extending to beyond-Horndeski constructions \cite{Kobayashi:2015gga}. 
The proof of this theorem postulates that the involved galileon functions $G_I$'s are 
non-singular, and that a specific quantity related to the tensor perturbation remains 
finite at the bounce point (see Eq. (10) of  \cite{Kobayashi:2016xpl}). Abandoning the 
first postulate allows for a stable non-singular bounce in the Horndeski class through 
galileon functions $G_I$'s that diverge at the bounce point, as it was shown in 
\cite{Ijjas:2016tpn}. In our bouncing solutions obtained in the present work, one can show 
that the second postulate is bypassed, and thus Kobayashi's no-go theorem is evaded. 
Hence, the scenario at hand is free of ghost 
instabilities and therefore we obtain a well behaved model in terms of perturbations. 
Since this issue has a separate interest, that is related to the full Horndeski theory 
and not only to its specific subclass of theories with weakly broken galileon invariance,
we are going to discuss it in detail in separate work \cite{uspreparation}.

\section{Conclusions}
\label{Conclusions}
 
 We have investigated the bounce and cyclicity realization in the framework of weakly 
broken galileon theories. In this subclass of modified gravity one introduces the notion 
of weakly broken galileon invariance, which characterizes the unique class 
of gravitational couplings that maximally preserve the defining symmetry. Hence, the 
curved-space remnants of the quantum properties of the galileon allow one to construct 
quasi de Sitter backgrounds that remain to a large extent insensitive to loop corrections 
\cite{Pirtskhalava:2015nla}.

We studied bouncing and cyclic solutions at the background level, reconstructing 
the potential that can give rise to a given bouncing or cyclic scale factor. Then, 
reversing the procedure, we considered suitable potential forms that can generate a 
bounce or cyclic behavior. Additionally, for a zero or non-zero given potential, we 
reconstructed the forms of the galileon functions that give rise to a bouncing solution.
Finally, we presented some analytical expressions for the 
requirements of bounce realization. As we showed, bounce and cyclicity can be easily 
realized in the framework of weakly broken galileon theories.

Having obtained the background bouncing solutions, we proceeded to a detailed 
investigation of the perturbations, which after crossing the bouncing point give rise to 
various observables, such as the scalar and tensor spectral indices and the 
tensor-to-scalar ratio. We calculated their values and we saw that the scenario at hand 
shares the disadvantage of all bouncing models, namely that it provides a large 
tensor-to-scalar ratio. Hence, we discussed about possible solutions, namely the 
possibility of introducing an additional light scalar which could significantly reduce 
the tensor-to-scalar ratio through the kinetic amplification of the isocurvature 
fluctuations. These features make the scenario at hand a good candidate for the 
description 
of the early universe.

%


\begin{thebibliography}{99}

  
   \bibitem{inflation}
  A.~H.~Guth,
 {\it{The Inflationary Universe: A Possible Solution To The Horizon And Flatness
  Problems}},
  Phys.\ Rev.\  D {\bf 23}, 347 (1981).
  
 
 
   
  
\bibitem{Borde:1993xh}
  A.~Borde and A.~Vilenkin,
 {\it{Eternal Inflation And The Initial Singularity}},
  Phys.\ Rev.\ Lett.\  {\bf 72}, 3305 (1994),
[\href{http://xxx.lanl.gov/abs/gr-qc/9312022}
{{\tt arXiv:gr-qc/9312022}}].


  
\bibitem{Mukhanov:1991zn}
  V.~F.~Mukhanov and R.~H.~Brandenberger,
 {\it{A Nonsingular universe}},
  Phys.\ Rev.\ Lett.\  {\bf 68}, 1969 (1992).
  
  
  
  
 
\bibitem{Nojiri:2006ri}
  S.~'i.~Nojiri and S.~D.~Odintsov,
  {\it{Introduction to modified gravity and gravitational alternative for
dark energy}},
  eConf C {\bf 0602061}, 06 (2006)
  [Int.\ J.\ Geom.\ Meth.\ Mod.\ Phys.\  {\bf 4}, 115 (2007)]
[\href{http://xxx.lanl.gov/abs/hep-th/0601213}
{{\tt arXiv:hep-th/0601213}}].

\bibitem{Capozziello:2011et}
  S.~Capozziello and M.~De Laurentis,
  {\it{Extended Theories of Gravity}},
  Phys.\ Rept.\  {\bf 509}, 167 (2011)
[\href{http://xxx.lanl.gov/abs/1108.6266}
{{\tt arXiv:1108.6266}}].



  
  
\bibitem{Veneziano:1991ek}
  G.~Veneziano,
 {\it{Scale Factor Duality For Classical And Quantum Strings}},
  Phys.\ Lett.\  B {\bf 265}, 287 (1991).
  
 

\bibitem{Khoury:2001wf}
  J.~Khoury, B.~A.~Ovrut, P.~J.~Steinhardt and N.~Turok,
 {\it{The ekpyrotic universe: Colliding branes and the origin of the hot
big bang}},
  Phys.\ Rev.\  D {\bf 64}, 123522
(2001),
[\href{http://xxx.lanl.gov/abs/hep-th/0103239}
{{\tt arXiv:hep-th/0103239}}].

 

 
 \bibitem{Khoury:2001bz}
  J.~Khoury, B.~A.~Ovrut, N.~Seiberg, P.~J.~Steinhardt and N.~Turok,
 {\it{From big crunch to big bang}},
  Phys.\ Rev.\  D {\bf 65}, 086007 (2002)
  [\href{http://xxx.lanl.gov/abs/hep-th/0108187}
{{\tt arXiv:hep-th/0108187}}].

  

 
  
  
\bibitem{Tirtho1}
  T.~Biswas, A.~Mazumdar and W.~Siegel,
 {\it{Bouncing universes in string-inspired gravity}},
  JCAP {\bf 0603}, 009 (2006),
[\href{http://xxx.lanl.gov/abs/hep-th/0508194}
{{\tt arXiv:hep-th/0508194}}].

 
  
  
  
\bibitem{Nojiri:2013ru} 
  S.~Nojiri and E.~N.~Saridakis,
   {\it{Phantom without ghost}},
  Astrophys.\ Space Sci.\  {\bf 347}, 221 (2013)
  [\href{http://xxx.lanl.gov/abs/1301.2686}
{{\tt arXiv:1301.2686}}].

\bibitem{Bamba:2013fha}
  K.~Bamba, A.~N.~Makarenko, A.~N.~Myagky, S.~Nojiri and S.~D.~Odintsov,
  {\it{Bounce cosmology from $F(R)$ gravity and $F(R)$ bigravity}},
  JCAP {\bf 1401} (2014) 008
    [\href{http://xxx.lanl.gov/abs/1309.3748}
{{\tt arXiv:1309.3748}}].


 
 
\bibitem{Nojiri:2014zqa} 
  S.~Nojiri and S.~D.~Odintsov,
   {\it{Mimetic $F(R)$ gravity: inflation, dark energy and bounce}},
  Mod.\ Phys.\ Lett.\ A {\bf 29}, no. 40, 1450211 (2014)
      [\href{http://xxx.lanl.gov/abs/1408.3561}
{{\tt arXiv:1408.3561}}].


 
  
\bibitem{Cai:2011tc}
  Y.~-F.~Cai, S.~-H.~Chen, J.~B.~Dent, S.~Dutta and E.~N.~Saridakis,
 {\it{Matter Bounce Cosmology with the f(T) Gravity}},
  Class.\ Quant.\ Grav.\  {\bf 28}, 215011 (2011),
[\href{http://xxx.lanl.gov/abs/1104.4349}
{{\tt arXiv:1104.4349}}].

  
  

 \bibitem{Shtanov:2002mb}
  Y.~Shtanov and V.~Sahni,
 {\it{Bouncing braneworlds}},
  Phys.\ Lett.\  B {\bf 557}, 1 (2003),
[\href{http://xxx.lanl.gov/abs/gr-qc/0208047}
{{\tt arXiv:gr-qc/0208047}}].


  
\bibitem{Saridakis:2007cf}
  E.~N.~Saridakis,
 {\it{Cyclic Universes from General Collisionless Braneworld Models}},
  Nucl.\ Phys.\ B {\bf 808}, 224 (2009),
[\href{http://xxx.lanl.gov/abs/0710.5269}
{{\tt arXiv:0710.5269}}].



\bibitem{Cai:2009in}
  Y.~F.~Cai and E.~N.~Saridakis,
 {\it{Non-singular cosmology in a model of non-relativistic gravity}},
  JCAP {\bf 0910}, 020 (2009),
[\href{http://xxx.lanl.gov/abs/0906.1789}
{{\tt arXiv:0906.1789}}].


\bibitem{Saridakis:2009bv}
  E.~N.~Saridakis,
 {\it{Horava-Lifshitz Dark Energy}},
  Eur.\ Phys.\ J.\ C {\bf 67}, 229 (2010),
[\href{http://xxx.lanl.gov/abs/0905.3532}
{{\tt arXiv:0905.3532}}]

  
\bibitem{Cai:2012ag} 
  Y.~F.~Cai, C.~Gao and E.~N.~Saridakis,
 {\it{Bounce and cyclic cosmology in extended nonlinear massive gravity}},
  JCAP {\bf 1210}, 048 (2012)
  [\href{http://xxx.lanl.gov/abs/1207.3786}
{{\tt arXiv:1207.3786}}]


 

\bibitem{Cai:2010zma}
  Y.~-F.~Cai and E.~N.~Saridakis,
 {\it{Cyclic cosmology from Lagrange-multiplier modified gravity}},
  Class.\ Quant.\ Grav.\  {\bf 28}, 035010 (2011),
[\href{http://xxx.lanl.gov/abs/1007.3204}
{{\tt arXiv:1007.3204}}].
  
 
  
   
  
  
\bibitem{Bojowald:2001xe}
  M.~Bojowald,
 {\it{Absence of singularity in loop quantum cosmology}},
  Phys.\ Rev.\ Lett.\  {\bf 86}, 5227 (2001),
[\href{http://xxx.lanl.gov/abs/gr-qc/0102069}
{{\tt arXiv:gr-qc/0102069}}].

 
 
  
  
\bibitem{Odintsov:2014gea} 
  S.~D.~Odintsov and V.~K.~Oikonomou,
 {\it{Matter Bounce Loop Quantum Cosmology from $F(R)$ Gravity}},
  Phys.\ Rev.\ D {\bf 90}, no. 12, 124083 (2014)
  [\href{http://xxx.lanl.gov/abs/1410.8183}
{{\tt arXiv:1410.8183}}].


 

\bibitem{Odintsov:2015uca} 
  S.~D.~Odintsov, V.~K.~Oikonomou and E.~N.~Saridakis,
   {\it{Superbounce and Loop Quantum Ekpyrotic Cosmologies from Modified Gravity: $F(R)$, 
$F(G)$ and $F(T)$ Theories}},
  Annals Phys.\  {\bf 363}, 141 (2015)
  [\href{http://xxx.lanl.gov/abs/1501.06591}
{{\tt arXiv:1501.06591}}].



\bibitem{Martin:2003sf}
  J.~Martin and P.~Peter,
 {\it{Parametric amplification of metric fluctuations through a bouncing
  phase}},
  Phys.\ Rev.\  D {\bf 68}, 103517 (2003),
[\href{http://xxx.lanl.gov/abs/hep-th/0307077}
{{\tt arXiv:hep-th/0307077}}].

    
\bibitem{Cai:2007qw}
  Y.~F.~Cai, T.~Qiu, Y.~S.~Piao, M.~Li and X.~Zhang,
 {\it{Bouncing Universe with Quintom Matter}},
  JHEP {\bf 0710}, 071 (2007),
[\href{http://xxx.lanl.gov/abs/0704.1090}
{{\tt arXiv:0704.1090}}].
  
  
  
  
\bibitem{Cai:2009zp}
  Y.~-F.~Cai, E.~N.~Saridakis, M.~R.~Setare and J.~-Q.~Xia,
 {\it{Quintom Cosmology: Theoretical implications and observations}},
  Phys.\ Rept.\  {\bf 493}, 1 (2010),
[\href{http://xxx.lanl.gov/abs/0909.2776}
{{\tt arXiv:0909.2776}}].

\bibitem{Nojiri:2015fia} 
  S.~Nojiri, S.~D.~Odintsov, V.~K.~Oikonomou and E.~N.~Saridakis,
  {\it{Singular cosmological evolution using canonical and ghost scalar fields}},
  JCAP {\bf 1509}, 044 (2015)
  [\href{http://xxx.lanl.gov/abs/1503.08443}
{{\tt arXiv:1503.08443}}].

 

  
  
\bibitem{Saridakis:2009jq} 
  E.~N.~Saridakis and J.~M.~Weller,
 {\it{A Quintom scenario with mixed kinetic terms}},
  Phys.\ Rev.\ D {\bf 81}, 123523 (2010)
  [\href{http://xxx.lanl.gov/abs/0912.5304}
{{\tt arXiv:0912.5304}}].



 
  
  

\bibitem{Saridakis:2009uk} 
  E.~N.~Saridakis and J.~Ward,
 {\it{Quintessence and phantom dark energy from ghost D-branes}},
  Phys.\ Rev.\ D {\bf 80}, 083003 (2009)
    [\href{http://xxx.lanl.gov/abs/0906.5135}
{{\tt arXiv:0906.5135}}].




 
 
  
  

\bibitem{tolman}
  R. C. Tolman,
  {\it{Relativity, Thermodynamics and Cosmology}},
  Oxford U. Press (1934).
  
\bibitem{Steinhardt:2001st}
  P.~J.~Steinhardt and N.~Turok,
 {\it{Cosmic evolution in a cyclic universe}},
  Phys.\ Rev.\  D {\bf 65}, 126003 (2002),
[\href{http://xxx.lanl.gov/abs/hep-th/0111098}
{{\tt arXiv:hep-th/0111098}}].



\bibitem{Steinhardt:2002ih} 
  P.~J.~Steinhardt and N.~Turok,
 {\it{A cyclic model of the universe}},
  Science {\bf 296}, 1436 (2002).
 
  

\bibitem{Lidsey:2004ef}
  J.~E.~Lidsey, D.~J.~Mulryne, N.~J.~Nunes and R.~Tavakol,
 {\it{Oscillatory universes in loop quantum cosmology and initial
conditions for inflation}},
  Phys.\ Rev.\  D {\bf 70}, 063521 (2004),
[\href{http://xxx.lanl.gov/abs/gr-qc/0406042}
{{\tt arXiv:gr-qc/0406042}}].


 
     
  
    
\bibitem{cyclic1}
  L.~Baum and P.~H.~Frampton,
 {\it{Turnaround in Cyclic Cosmology}},
  Phys.\ Rev.\ Lett.\  {\bf 98}, 071301 (2007),
[\href{http://xxx.lanl.gov/abs/hep-th/0610213}
{{\tt arXiv:hep-th/0610213}}].
  

\bibitem{Nojiri:2011kd} 
  S.~Nojiri, S.~D.~Odintsov and D.~Saez-Gomez,
 {\it{Cyclic, ekpyrotic and little rip universe in modified gravity}},
  AIP Conf.\ Proc.\  {\bf 1458}, 207 (2011)
  [\href{http://xxx.lanl.gov/abs/1108.0767}
{{\tt arXiv:1108.0767}}].

 
   

\bibitem{Novello:2008ra}
  M.~Novello and S.~E.~P.~Bergliaffa,
 {\it{Bouncing Cosmologies}},
  Phys.\ Rept.\  {\bf 463}, 127 (2008),
[\href{http://xxx.lanl.gov/abs/0802.1634}
{{\tt arXiv:0802.1634}}].

 
 

\bibitem{Finelli:2001sr} 
  F.~Finelli and R.~Brandenberger,
 {\it{On the generation of a scale invariant spectrum of adiabatic fluctuations in 
cosmological models with a contracting phase}},
  Phys.\ Rev.\ D {\bf 65}, 103522 (2002)
    [\href{http://xxx.lanl.gov/abs/hep-th/0112249}
{{\tt arXiv:hep-th/0112249}}].
 
  
 
 
\bibitem{Cai:2009fn}
  Y.~-F.~Cai, W.~Xue, R.~Brandenberger and X.~Zhang,
 {\it{Non-Gaussianity in a Matter Bounce}},
  JCAP {\bf 0905}, 011 (2009),
[\href{http://xxx.lanl.gov/abs/0903.0631}
{{\tt arXiv:0903.0631}}].

  
 
 
 
 
 
 
 
 
 
 
 
 

\bibitem{Nicolis:2008in}
  A.~Nicolis, R.~Rattazzi and E.~Trincherini,
{\it{The galileon as a local modification of gravity}},
  Phys.\ Rev.\  D {\bf 79}, 064036 (2009)
[\href{http://xxx.lanl.gov/abs/0811.2197}
{{\tt arXiv:0811.2197}}].


\bibitem{Deffayet:2009wt}
C.~Deffayet, G.~Esposito-Farese, and A.~Vikman, {\it{Covariant galileon}},
 Phys.\ Rev.\ D {\bf 79}, 084003 (2009)
[\href{http://xxx.lanl.gov/abs/0901.1314}
{{\tt arXiv:0901.1314}}].

  
  

\bibitem{Deffayet:2009mn}
C.~Deffayet, S.~Deser, and G.~Esposito-Farese, {\it{Generalized galileons:
All scalar models whose curved background extensions maintain second-order
field equations and stress-tensors}},
  Phys.\ Rev.\ D {\bf 80}, 064015 (2009)
[\href{http://xxx.lanl.gov/abs/0906.1967}
{{\tt arXiv:0906.1967}}].


 
 
  
\bibitem{Horndeski:1974wa}
  G.~W.~Horndeski,
  {\it{Second-order scalar-tensor field equations in a four-dimensional space}},
  Int.\ J.\ Theor.\ Phys.\  {\bf 10} (1974) 363.

\bibitem{Vainshtein:1972sx}
  A.~I.~Vainshtein,
  {\it{To the problem of nonvanishing gravitation mass}},
  Phys.\ Lett.\ B {\bf 39}, 393 (1972).


   
  
  
\bibitem{Silva:2009km}
  F.~P. Silva and K.~Koyama,
  {\it{Self-Accelerating Universe in galileon Cosmology}},
  Phys.\ Rev.\ D {\bf 80}, 121301 (2009)
[\href{http://xxx.lanl.gov/abs/0909.4538}
{{\tt arXiv:0909.4538}}].
 
\bibitem{DeFelice:2010pv}
  A.~De Felice and S.~Tsujikawa,
  {\it{Cosmology of a covariant galileon field}},
  Phys.\ Rev.\ Lett.\  {\bf 105}, 111301 (2010)
[\href{http://xxx.lanl.gov/abs/1007.2700}
{{\tt arXiv:1007.2700}}].


 
 

\bibitem{Gannouji:2010au}
  R.~Gannouji and M.~Sami,
  {\it{galileon gravity and its relevance to late time cosmic
acceleration}},
  Phys.\ Rev.\ D {\bf 82}, 024011 (2010)
[\href{http://xxx.lanl.gov/abs/1004.2808}
{{\tt arXiv:1004.2808}}].
 

\bibitem{Tretyakov:2012zz}
  P.~Tretyakov,
       {\it{Scaling solutions in galileon cosmology}},
  Grav.\ Cosmol.\  {\bf 18}, 93 (2012).
 
 
\bibitem{Leon:2012mt} 
  G.~Leon and E.~N.~Saridakis,
    {\it{Dynamical analysis of generalized Galileon cosmology}},
  JCAP {\bf 1303}, 025 (2013)
  [\href{http://xxx.lanl.gov/abs/1211.3088}
{{\tt arXiv:1211.3088}}].
 
  


\bibitem{Creminelli:2010ba}
  P.~Creminelli, A.~Nicolis and E.~Trincherini,
  {\it{Galilean Genesis: An Alternative to inflation}},
  JCAP {\bf 1011}, 021 (2010)
[\href{http://xxx.lanl.gov/abs/1007.0027}
{{\tt arXiv:1007.0027}}].




\bibitem{Kobayashi:2010cm}
  T.~Kobayashi, M.~Yamaguchi and J.~'i.~Yokoyama,
  {\it{G-inflation: Inflation driven by the galileon field}},'
  Phys.\ Rev.\ Lett.\  {\bf 105}, 231302 (2010)
[\href{http://xxx.lanl.gov/abs/1008.0603}
{{\tt arXiv:1008.0603}}].


 
 

\bibitem{Ohashi:2012wf}
  J.~Ohashi and S.~Tsujikawa,
       {\it{Potential-driven galileon inflation}},
[\href{http://xxx.lanl.gov/abs/1207.4879}
{{\tt arXiv:1207.4879}}].
 

\bibitem{Mizuno:2010ag}
  S.~Mizuno and K.~Koyama,
  {\it{Primordial non-Gaussianity from the DBI galileons}},
  Phys.\ Rev.\ D {\bf 82}, 103518 (2010)
[\href{http://xxx.lanl.gov/abs/1009.0677}
{{\tt arXiv:1009.0677}}].

  
\bibitem{Gao:2011qe}
  X.~Gao and D.~A.~Steer,
  {\it{Inflation and primordial non-Gaussianities of 'generalized
galileons'}},
  JCAP {\bf 1112}, 019 (2011)
[\href{http://xxx.lanl.gov/abs/1107.2642}
{{\tt arXiv:1107.2642}}].






\bibitem{RenauxPetel:2011uk}
  S.~Renaux-Petel, S.~Mizuno and K.~Koyama,
  {\it{Primordial fluctuations and non-Gaussianities from multifield DBI
galileon inflation}},
  JCAP {\bf 1111}, 042 (2011)
[\href{http://xxx.lanl.gov/abs/1108.0305}
{{\tt arXiv:1108.0305}}].


 
 
 



\bibitem{Kobayashi:2009wr}
  T.~Kobayashi, H.~Tashiro and D.~Suzuki,
  {\it{Evolution of linear cosmological perturbations and its observational
implications in galileon-type modified gravity}},
  Phys.\ Rev.\ D {\bf 81}, 063513 (2010)
[\href{http://xxx.lanl.gov/abs/0912.4641}
{{\tt arXiv:0912.4641}}].




\bibitem{DeFelice:2010as}
  A.~De Felice, R.~Kase and S.~Tsujikawa,
  {\it{Matter perturbations in galileon cosmology}},
  Phys.\ Rev.\ D {\bf 83}, 043515 (2011)
[\href{http://xxx.lanl.gov/abs/1011.6132}
{{\tt arXiv:1011.6132}}].



 
 

\bibitem{Barreira:2012kk}
  A.~Barreira, B.~Li, C.~Baugh and S.~Pascoli,
       {\it{Linear perturbations in galileon gravity models}},
  Phys.\ Rev.\ D {\bf 86}, 124016 (2012)
[\href{http://xxx.lanl.gov/abs/1208.0600}
{{\tt arXiv:1208.0600}}].




\bibitem{Ali:2010gr}
  A.~Ali, R.~Gannouji and M.~Sami,
  {\it{Modified gravity a la galileon: Late time cosmic acceleration and
observational constraints}},
  Phys.\ Rev.\ D {\bf 82}, 103015 (2010)
[\href{http://xxx.lanl.gov/abs/1008.1588}
{{\tt arXiv:1008.1588}}].

 

\bibitem{Appleby:2012ba}
  S.~A.~Appleby and E.~V.~Linder,
       {\it{Trial of galileon gravity by cosmological expansion and growth
observations}},
  JCAP {\bf 1208}, 026 (2012)
[\href{http://xxx.lanl.gov/abs/1204.4314}
{{\tt arXiv:1204.4314}}].

 


\bibitem{Iorio:2012pv}
  L.~Iorio,
  {\it{Constraints on galileon-induced precessions from solar system
orbital motions}},
  JCAP {\bf 1207}, 001 (2012)
[\href{http://xxx.lanl.gov/abs/1204.0745}
{{\tt arXiv:1204.0745}}].

 

  
   
  
\bibitem{Pirtskhalava:2015nla} 
  D.~Pirtskhalava, L.~Santoni, E.~Trincherini and F.~Vernizzi, 
  {\it{Weakly Broken 
galileon 
Symmetry}},
JCAP {\bf 1509} 007 (2015) 
[\href{http://xxx.lanl.gov/abs/1505.00007}
{{\tt arXiv:1505.00007}}].


 

 
 
\bibitem{Qiu:2011cy}
  T.~Qiu, J.~Evslin, Y.~-F.~Cai, M.~Li and X.~Zhang,
  {\it{Bouncing galileon Cosmologies}},
  JCAP {\bf 1110}, 036 (2011)
[\href{http://xxx.lanl.gov/abs/1108.0593}
{{\tt arXiv:1108.0593}}].





\bibitem{Easson:2011zy}
  D.~A.~Easson, I.~Sawicki and A.~Vikman,
  {\it{G-Bounce}},
  JCAP {\bf 1111}, 021 (2011)
[\href{http://xxx.lanl.gov/abs/1109.1047}
{{\tt arXiv:1109.1047}}].

\bibitem{Cai:2012va} 
  Y.~-F.~Cai, D.~A.~Easson and R.~Brandenberger,
  {\it{Towards a Nonsingular Bouncing Cosmology}},
  JCAP {\bf 1208}, 020 (2012)
[\href{http://xxx.lanl.gov/abs/1206.2382}
{{\tt arXiv:1206.2382}}].
  
\bibitem{Qiu:2013eoa} 
  T.~Qiu, X.~Gao and E.~N.~Saridakis,
 {\it{Towards anisotropy-free and nonsingular bounce cosmology with scale-invariant 
perturbations}},
  Phys.\ Rev.\ D {\bf 88}, no. 4, 043525 (2013)
  [\href{http://xxx.lanl.gov/abs/1303.2372}
{{\tt arXiv:1303.2372}}].

\bibitem{Belinsky:1970ew} 
  V.~A.~Belinsky, I.~M.~Khalatnikov and E.~M.~Lifshitz,
 {\it{Oscillatory approach to a singular point in the relativistic cosmology}},
  Adv.\ Phys.\  {\bf 19}, 525 (1970).


\bibitem{Cai:2013vm} 
  Y.~F.~Cai, R.~Brandenberger and P.~Peter,
 {\it{Anisotropy in a Nonsingular Bounce}},
  Class.\ Quant.\ Grav.\  {\bf 30}, 075019 (2013),
    [\href{http://xxx.lanl.gov/abs/1301.4703}
{{\tt arXiv:1301.4703}}].

\bibitem{Cai:2013kja} 
  Y.~F.~Cai, E.~McDonough, F.~Duplessis and R.~H.~Brandenberger,
 {\it{Two Field Matter Bounce Cosmology}},
  JCAP {\bf 1310}, 024 (2013),
      [\href{http://xxx.lanl.gov/abs/1305.5259}
{{\tt arXiv:1305.5259}}].

\bibitem{Cai:2014xxa} 
  Y.~F.~Cai, J.~Quintin, E.~N.~Saridakis and E.~Wilson-Ewing,
 {\it{Nonsingular bouncing cosmologies in light of BICEP2}},
  JCAP {\bf 1407}, 033 (2014),
        [\href{http://xxx.lanl.gov/abs/1404.4364}
{{\tt arXiv:1404.4364}}].


 
\bibitem{Cai:2014bea} 
  Y.~F.~Cai,
 {\it{Exploring Bouncing Cosmologies with Cosmological Surveys}},
  Sci.\ China Phys.\ Mech.\ Astron.\  {\bf 57}, 1414 (2014),
          [\href{http://xxx.lanl.gov/abs/1405.1369}
{{\tt arXiv:1405.1369}}].

\bibitem{Brandenberger:2009ic} 
  R.~H.~Brandenberger,
  {\it{Processing of Cosmological Perturbations in a Cyclic Cosmology}},
  Phys.\ Rev.\ D {\bf 80}, 023535 (2009)
            [\href{http://xxx.lanl.gov/abs/0905.1514}
{{\tt arXiv:0905.1514}}].

\bibitem{Piao:2010cf} 
  Y.~S.~Piao,
  {\it{Design of a Cyclic Multiverse}},
  Phys.\ Lett.\ B {\bf 691}, 225 (2010)
              [\href{http://xxx.lanl.gov/abs/1001.0631}
{{\tt arXiv:1001.0631}}].

 

 

\bibitem{Starobinsky:1979ty}
  A.~A.~Starobinsky,
  {\it{Spectrum of relict gravitational radiation and the early state of the
  universe}},
  JETP Lett.\ \textbf{30} (1979) 682
  [Pisma Zh.\ Eksp.\ Teor.\ Fiz.\ \textbf{30} (1979) 719].



\bibitem{Battarra:2014tga} 
  L.~Battarra, M.~Koehn, J.~L.~Lehners and B.~A.~Ovrut,
 {\it{Cosmological Perturbations Through a Non-Singular Ghost-Condensate/Galileon 
Bounce}},
  JCAP {\bf 1407}, 007 (2014)
      [\href{http://xxx.lanl.gov/abs/1404.5067}
{{\tt arXiv:1404.5067}}].

\bibitem{Quintin:2015rta} 
  J.~Quintin, Z.~Sherkatghanad, Y.~F.~Cai and R.~H.~Brandenberger,
 {\it{Evolution of cosmological perturbations and the production of non-Gaussianities 
through a nonsingular bounce: Indications for a no-go theorem in single field matter 
bounce cosmologies}},
  Phys.\ Rev.\ D {\bf 92}, no. 6, 063532 (2015)
        [\href{http://xxx.lanl.gov/abs/1508.04141}
{{\tt arXiv:1508.04141}}].


\bibitem{Koehn:2015vvy} 
  M.~Koehn, J.~L.~Lehners and B.~Ovrut,
  {\it{Nonsingular bouncing cosmology: Consistency of the effective description}},
  Phys.\ Rev.\ D {\bf 93}, no. 10, 103501 (2016)
          [\href{http://xxx.lanl.gov/abs/1512.03807}
{{\tt arXiv:1512.03807}}].
 
 
  
  
  
   \bibitem{Piazza:2013coa}
  F.~Piazza and F.~Vernizzi,
 {\it{Effective Field Theory of Cosmological Perturbations}},
  Class.\ Quant.\ Grav.\  {\bf 30}, 214007 (2013)
  [\href{http://xxx.lanl.gov/abs/1307.4350}
{{\tt arXiv:1307.4350}}].

  
  
 


  \bibitem{Gleyzes:2013ooa}
  J.~Gleyzes, D.~Langlois, F.~Piazza and F.~Vernizzi,
   {\it{Essential Building Blocks of Dark Energy}},
  JCAP {\bf 1308}, 025 (2013)
  [\href{http://xxx.lanl.gov/abs/1304.4840}
{{\tt arXiv:1304.4840}}].


  



\bibitem{Maldacena:2002vr}
  J.~M.~Maldacena,
  {\it{Non-Gaussian features of primordial fluctuations in single field inflationary 
models}},
  JHEP {\bf 0305}, 013 (2003)
    [\href{http://xxx.lanl.gov/abs/astro-ph/0210603}
{{\tt arXiv:astro-ph/0210603}}].


\bibitem{Allen:1985ux} 
  B.~Allen,
  {\it{Vacuum States in de Sitter Space}},
  Phys.\ Rev.\ D {\bf 32}, 3136 (1985).

  
  
\bibitem{Planck:2015xua} 
  P.~A.~R.~Ade {\it et al.}  [Planck Collaboration],
  {\it{Planck 2015 results. XIII. Cosmological parameters}},
      [\href{http://xxx.lanl.gov/abs/1502.01589}
{{\tt arXiv:1502.01589}}].


  


\bibitem{Array:2015xqh} 
  P.~A.~R.~Ade {\it et al.} [BICEP2 and Keck Array Collaborations],
   {\it{Improved Constraints on Cosmology and Foregrounds from BICEP2 and 
  Keck Array Cosmic Microwave Background Data with Inclusion 
  of 95 GHz Band}},
  Phys.\ Rev.\ Lett.\  {\bf 116}, 031302 (2016)
        [\href{http://xxx.lanl.gov/abs/1510.09217}
{{\tt arXiv:1510.09217}}].


\bibitem{Cai:2011zx}
  Y.~F.~Cai, R.~Brandenberger and X.~Zhang,
    {\it{The Matter Bounce Curvaton Scenario}},
  JCAP {\bf 1103}, 003 (2011)
          [\href{http://xxx.lanl.gov/abs/1101.0822}
{{\tt arXiv:1101.0822}}].

\bibitem{Kobayashi:2016xpl} 
  T.~Kobayashi,
   {\it{Generic instabilities of nonsingular cosmologies in Horndeski theory: A no-go 
theorem}},
  Phys.\ Rev.\ D {\bf 94}, no. 4, 043511 (2016)
            [\href{http://xxx.lanl.gov/abs/1606.05831}
{{\tt arXiv:1606.05831}}].
 

\bibitem{Akama:2017jsa} 
  S.~Akama and T.~Kobayashi,
   {\it{Generalized multi-Galileons, covariantized new terms, and the no-go theorem for 
non-singular cosmologies}},
            [\href{http://xxx.lanl.gov/abs/1701.02926}
{{\tt arXiv:1701.02926}}].
 
\bibitem{Kobayashi:2015gga} 
  T.~Kobayashi, M.~Yamaguchi and J.~Yokoyama,
   {\it{Galilean Creation of the Inflationary Universe}},
  JCAP {\bf 1507}, no. 07, 017 (2015)
      [\href{http://xxx.lanl.gov/abs/1504.05710}
{{\tt arXiv:1504.05710}}].

 
  
  
\bibitem{Ijjas:2016tpn} 
  A.~Ijjas and P.~J.~Steinhardt,
   {\it{Classically stable nonsingular cosmological bounces}},
  Phys.\ Rev.\ Lett.\  {\bf 117}, no. 12, 121304 (2016)
    [\href{http://xxx.lanl.gov/abs/1606.08880}
{{\tt arXiv:1606.08880}}].

 
 
\bibitem{uspreparation} 
  Shreya Banerjee, Y.~F.~Cai, Emmanuel N. Saridakis and Youping Wan,
   {\it{in preparation}}.
 
  
\end{thebibliography}
\end{document}